\documentclass[aps,pre,showpacs,amsmath,amssymb,superscriptaddress,reprint,10pt,longbibliography]{revtex4-1}
\usepackage{bm}
\usepackage{graphicx}
\usepackage{xcolor}
\usepackage{url}
\usepackage[normalem]{ulem}
\usepackage{times}
\usepackage{bbm}
\usepackage{mathrsfs} 
\usepackage{comment}
\usepackage{dsfont} 
\usepackage{multirow} 
\usepackage{ulem}

\usepackage[colorlinks=true,breaklinks=true,allcolors=blue]{hyperref}
\usepackage{url}

\newcommand\NN{{\mathds{N}}}

\newcommand\CC{{\mathds{C}}}
\newcommand\id{{\mathds{1}}}

\newcommand{\mvec}[1]{\boldsymbol #1}

\newcommand{\couple}{\mathscr{U}}
\newcommand{\Id}{\mathds{1}}
\newcommand{\wmcf}{\mathscr{C}}
\DeclareMathOperator{\myIm}{\mathrm{Im}}

\newcommand{\ket}[1]{\left \vert #1 \right >}
\newcommand{\bra}[1]{\left < #1 \right \vert}
\newcommand{\matel}[3]{ \displaystyle \left\langle #1 \right \vert #2 \left\vert  #3 \right\rangle }
\newcommand{\expval}[1]{\left\langle #1 \right\rangle}
\newcommand{\innprod}[2]{ \displaystyle \left\langle #1 \vert #2 \right\rangle }
\newcommand{\impart}[1]{\text{Im}\left[ #1 \right]}
\newcommand{\realpart}[1]{\text{Re}\left[ #1 \right]}

\newcommand{\absval}[1]{\left\vert #1 \right\vert}
\newcommand{\projector}[1]{\ket{#1} \bra{#1}}
\DeclareMathOperator{\estDev}{\epsilon_{\text{tot}}}
\DeclareMathOperator{\eSyst}{\epsilon_{\text{sys}}}
\DeclareMathOperator{\eFS}{\epsilon_{\text{stat}}}

\begin{document}


\title{Noninvasive Measurement of Dynamic Correlation Functions}  

\author{Philipp Uhrich} 
\affiliation{National Institute for Theoretical Physics (NITheP), Stellenbosch 7600, South Africa} 
\affiliation{Institute of Theoretical Physics,  Department of Physics, University of Stellenbosch, Stellenbosch 7600, South Africa}

\author{Salvatore Castrignano}
\affiliation{Max-Planck-Institut f\"ur Kernphysik, Saupfercheckweg 1, 69117 Heidelberg, Germany}

\author{Hermann Uys}
\affiliation{Institute of Theoretical Physics,  Department of Physics, University of Stellenbosch, Stellenbosch 7600, South Africa}
\affiliation{Council for Scientific and Industrial Research, National Laser Centre, Pretoria, Brummeria, 0184, South Africa}

\author{Michael Kastner} 
\email{kastner@sun.ac.za} 
\affiliation{National Institute for Theoretical Physics (NITheP), Stellenbosch 7600, South Africa} 
\affiliation{Institute of Theoretical Physics,  Department of Physics, University of Stellenbosch, Stellenbosch 7600, South Africa}

\date{\today}

\begin{abstract}
The measurement of dynamic correlation functions of quantum systems is complicated by measurement backaction. To facilitate such measurements we introduce a protocol, based on weak ancilla--system couplings, that is applicable to arbitrary (pseudo)spin systems and arbitrary equilibrium or nonequilibrium initial states. Different choices of the coupling operator give access to the real and imaginary parts of the dynamic correlation function. This protocol reduces disturbances due to the early time measurements to a minimum, and we quantify the deviation of the measured correlation functions from the theoretical, unitarily-evolved ones. Implementations of the protocol in trapped ions and other experimental platforms are discussed. For spin-$1/2$ models and single-site observables we prove that measurement backaction can be avoided altogether, allowing for the use of ancilla-free protocols.
\end{abstract}


\maketitle 

\section{Introduction}
Dynamic correlation functions such as $\left\langle O_1(t_1)O_2(t_2)\right\rangle$ relate the values of some observable $O_1$ at an early time $t_1$ to the value of another observable $O_2$ at a later time $t_2$. They play an important role in many theoretical approaches, including fluctuation-dissipation theorems and the Kubo formula \cite{Kubo57}, optical coherence \cite{Glauber63}, glassy dynamics and aging \cite{SciollaPolettiKollath15}, and many more.

In a classical (non-quantum mechanical) system, a straight\-forward---at least in principle---protocol for determining dynamic correlations consists of measuring the observable $O_1$ at time $t_1$ and correlating the outcome with the measured value of $O_2$ at time $t_2$. In a quantum mechanical system, however, such a naive approach is in general thwarted by the measurement backaction, i.e., by the disturbing effect that a measurement of $O_1$ at the earlier time $t_1$ has, due to the collapse of the wave function, on the subsequent time evolution \cite{MakhlinSchoenShnirman00,DiLorenzo_etal06,Oehri_etal14}. As a result of this disturbance, correlating the outcomes of measuring $O_1$ at time $t_1$ with that of $O_2$ at $t_2$ does not yield the desired dynamic correlation function.
As an example, consider two spin-$1/2$ degrees of freedom, initially in a product state $\ket{\psi} = (\alpha \ket{+} + \beta \ket{-})\otimes(\alpha \ket{+} + \beta \ket{-})$ with $\alpha, \beta\in\CC$, where $\ket{+}$ and $\ket{-}$ denote eigenstates of the Pauli operator $\sigma^z$ with eigenvalues $+1$ and $-1$, respectively. Assume the dynamics of the two spins to be governed by the Hamiltonian $H=\sigma^x \otimes \sigma^x$. For this scenario, a simple calculation (reported in Appendix \ref{app: naive_prot_eg}) shows that
\begin{multline}\label{e:corr_example}
\matel{\psi}{\sigma^z (0) \otimes \sigma^z(t)}{\psi} = \cos(2t) \left( \absval{\alpha}^{2} - \absval{\beta}^{2} \right)^2 \\
-i \sin(2t) \left( \alpha^* \beta - \alpha \beta^* \right)^2
\end{multline}
in units where $\hbar=1$. The above mentioned naive protocol for obtaining dynamic correlations by projective measurements, however, fails to reproduce this result; see Appendix \ref{app: naive_prot_eg}.

Due to this failure, measurements of dynamic correlations of quantum systems can be challenging. An interesting scheme, based on Ramsey interferometry and spin-shelving, for probing thermal equilibrium values of dynamic correlations has been put forward by Knap {\em et al}.\ \cite{Knap_etal13}. This scheme requires certain symmetries of the Hamiltonian, and gives access only to the imaginary part of certain components of dynamic correlations, and to the real part of other components. Another protocol for measuring dynamic correlations, which is due to Romero-Isart {\em et al.}\ \cite{RomeroIsart_etal12}, proposes to weakly couple photons to ultracold atoms in an optical lattice, and store the information imprinted on the photons in a quantum memory. Reading out the correlations between the system and the quantum memory at a later time then gives access to the real part of the dynamic correlation function. Here we introduce a different method, also based on weak system--ancilla coupling, which does not require quantum memories, allows one to measure real as well as imaginary parts of two-time dynamic correlation functions, and applies to arbitrary quantum spin systems and arbitrary equilibrium or nonequilibrium initial states. The setting we have in mind is a spatially extended system, and for simplicity we focus on lattice models. We consider dynamic correlation functions $\left\langle O_i(t_1)O_j(t_2)\right\rangle$, where the observables $O_i$ and $O_j$ act nontrivially only on lattice sites $i$ and $j$. 

Our first main result is a protocol for determining dynamic correlations $\left\langle O_i(t_1)O_j(t_2)\right\rangle$ by means of noninvasive measurements. Noninvasive measurements have been around for some time and under various names, including nonprojective, generalized, unsharp, or weak measurements \footnote{ 
We refrain from using the terminology ``weak measurement'' in order to avoid confusion with the concept of a (postselected) weak value, which plays no role in our protocol.
}, and these names are used for slightly different concepts in some works, and interchangeably in others; see \cite{Svensson13} for an introduction. The key idea of a noninvasive measurement is simple: Instead of making a measurement on the quantum system directly, a quantum mechanical ancilla is weakly coupled to it for a short period of time. By subsequently making a measurement on the ancilla, some information about the quantum system of interest is retrieved, but a full projection of the system's state onto an eigenstate of the measured observable is avoided. Noninvasive measurements play an important role in continuous measurements \cite{BraginskyKhalili,JacobsSteck06} and quantum control \cite{WisemanMilburn}, and they have also been used for quantum state estimation \cite{DasArvind14}.

The basic idea behind the noninvasive measurement protocol, introduced in detail in Sec.~\ref{sec:NIM_protocol}, is simple: The system of interest is let to evolve unitarily until the time $t_1$. At that time, an ancillary quantum system is weakly coupled to lattice site $i$ for a short period of time, after which a small amount of information about the system is retrieved by performing a projective measurement on the ancilla. Then, with the ancilla decoupled, the system is evolved unitarily until time $t_2$, at which $O_j$ is measured projectively. The novel technical finding here is to identify specific choices of the weak-coupling unitaries that give access to the real, respectively imaginary, parts of the correlation function. We show in Sec.~\ref{sec:estimator_construction} that the information obtained through multiple repetitions of this protocol can, for sufficiently weak system--ancilla coupling, be assembled to construct a faithful estimator of $\left\langle O_i(t_1)O_j(t_2)\right\rangle$. In Sec.~\ref{sec:numeric_imp} we characterize the performance of the noninvasive measurement protocol by deriving error bounds on the estimators for the dynamic correlation functions. These error bounds allow us to determine the optimal weak-coupling strength for a given number of repetitions of the protocol, which enables us to simultaneously minimize statistical and systematic errors. In Sec.~\ref{s:generalizations} we discuss generalizations of the noninvasive measurement protocol. The first is based on deferred measurements where information about the system at an early time $t_1$ is stored in an ancilla but read out not before $t_2$. We show in Appendix~\ref{s:deferred} that deferral yields no further reduction of the backaction. A second generalization uses multiple noninvasive measurements at times $t_1$, $t_2$, $t_3$, but it turns out that such a scheme is not advantageous. Our noninvasive measurement protocol is versatile, but also experimentally demanding in that multiple repetitions of the experiment are required, and a high degree of control is needed, in particular the possibility to couple and decouple an ancilla to the system. We discuss in Sec.~\ref{sec:exp_imp} an implementation of the protocol with ions in a linear Paul trap where the required steps can be realized with available experimental technology.

Our second main result, reported in Sec.~\ref{s:DPMP}, is specific to spin-$1/2$ systems: we prove for general spin-$1/2$ Hamiltonians that the real part of $\left\langle \sigma^a_i(t_1)\sigma^b_j(t_2)\right\rangle$  with $a,b \in \lbrace x,y,z \rbrace$ is not affected by measurement backaction. Hence, fully projective measurements can be used at times $t_1$ and $t_2$. This does not mean that no collapse of the wavefunction takes place, only that its effect precisely cancels out in the real part. The imaginary part of $\left\langle \sigma^a_i(t_1)\sigma^b_j(t_2)\right\rangle$ can be obtained by a different kind of measurement protocol, reported in Sec.~\ref{s:RPMP}, based on a local rotation of the spin at site $i$ at the early time $t_1$. Combining these two protocols, dynamic correlation functions can be obtained without the complications that arise from the use of an ancilla, while strictly avoiding any kind of backaction effects. From an experimental point of view this finding leads to a substantial simplification when dealing with spin-$1/2$ systems. 

\section{Noninvasive measurement protocol}
\label{sec:NIM_protocol}

All protocols are derived and stated in the language of lattice spin systems with spin quantum number $s\in \NN/2$, but generalizations to continuum systems should be possible.
Our aim is to estimate dynamic correlations 
\begin{align}\label{e:C_th}
C(t_1,t_2) &= \left\langle S_i^a(t_1)S_j^b(t_2)\right\rangle \nonumber \\
&= \matel{\psi}{ e^{i H t_1} S_{i}^{a}e^{-i H t_1}e^{i H t_2} S_j^{b}e^{-i H t_2}}{\psi},
\end{align}
where $S_i^a$ denotes the $a$-component of a spin-$s$ operator at lattice site $i$, with $a\in\{x,y,z\}$. For notational simplicity the Hamiltonian $H$ is assumed to be time-independent, but this constraint can be released. $\ket{\psi}$ is the initial system state at time $t=0$. Generalizations to correlations at more than two times and/or more than two lattice sites are possible and straightforward.

The possible outcomes of a projective measurement of either spin observable in \eqref{e:C_th} are $m_a,m_b \in \mathscr{S} = \lbrace s, s-1, \ldots, -s+1,-s\rbrace$. Performing such a measurement at times $t_1$ and $t_2$, the correlations between the early and the late measurement is given by
\begin{equation}\label{e:C_proj}
\mathscr{C}^{\text{proj}} = \sum_{m_a,m_b \in \mathscr{S}} m_a m_b P_{m_a m_b}  ,
\end{equation}
where $P_{m_a,m_b}$ denotes the joint probability to projectively measure eigenvalue $m_a$ and $m_b$ at times $t_1$ and $t_2$, respectively. Measuring such a correlation function by means of projective measurements at times $t_1$ and $t_2$ suffers from two difficulties (see Appendix \ref{app: naive_prot_eg} for a worked example). Firstly, the expectation value in \eqref{e:C_th} is in general complex, and therefore cannot be directly described by the real (non-complex) measurement outcomes and the corresponding probabilities as in \eqref{e:C_proj}. Secondly, as alluded to in the introduction, a projective measurement at the early time $t_1$ disturbs the unitary dynamics, and \eqref{e:C_th} and \eqref{e:C_proj} therefore differ in general. The following protocol, based on noninvasive measurements, successfully deals with both these difficulties.

For the noninvasive measurement at time $t_1$, the protocol makes use of an ancillary spin-$s$ degree of freedom. The total Hilbert space is therefore
\begin{equation}
\mathscr{H} = \mathscr{H}_{\text{A}} \otimes \mathscr{H}_{\text{S}},
\end{equation}
where the ancilla Hilbert space is $\mathscr{H}_{\text{A}}=\CC^{2s+1}$ and the Hilbert space for a system of $N$ spin-$s$ degrees of freedom is $\mathscr{H}_{\text{S}}=\left(\CC^{2s+1}\right)^{\otimes N}$. The system Hamiltonian $H=\id_{\text{A}}\otimes H_{\text{S}}$, which is responsible for the unitary evolution in the dynamic correlation function \eqref{e:C_th}, acts nontrivially on $\mathscr{H}_{\text{S}}$ only. The motivation for introducing the ancilla is that, by weakly coupling the ancilla to the system by means of a Hamiltonian that acts on the total Hilbert space $\mathscr{H}$, information about the system can be extracted by projectively measuring the ancilla, without causing a complete collapse of the system's wave function.

The noninvasive measurement protocol consists of the following steps.

\paragraph{Initial state preparation.}
We assume ancilla and system to initially be in a product state,
\begin{equation}\label{e:phi_psi}
\ket{\Psi}=\ket{\phi}\otimes\ket{\psi}\equiv\ket{\phi,\psi}.
\end{equation}
While the system initial state $\ket{\psi}$ is arbitrary (and determined by the physical situation under investigation), we will determine the optimal choice of the ancilla initial state $\ket{\phi}$ in \eqref{e:phi}.

\paragraph{Time evolution until time $t_1$.}
Time-evolve the initial state $\ket{\Psi}$ up to the time $t_1$ with the system Hamiltonian $H$,
\begin{equation}
\ket{\Psi(t_1)}=\ket{\phi}\otimes e^{-i H_s t_1}\ket{\psi}\equiv\ket{\phi,\psi(t_1)} .
\end{equation}
The ancilla state $\ket{\phi}$ remains unaffected.

\paragraph{Weak coupling of ancilla and system site $i$.\label{para:couple}} 
Time evolution of $\ket{\Psi(t_1)}$ with a coupling Hamiltonian $B\otimes A_i$ has the effect of generating entanglement between ancilla and system. The operator $A_i$ is chosen such as to act nontrivially only on the spin at lattice site $i$ for which, according to \eqref{e:C_th}, correlations at time $t_1$ are to be determined. This choice is expected to be most conducive towards our goal of imprinting information specifically about the state of the spin at site $i$ onto the ancilla. We assume that the corresponding time evolution operator
\begin{equation}\label{eq: coupling_linear}
 \couple(\lambda) = \exp(-i \lambda B \otimes A_i) \simeq \id - i \lambda B \otimes A_i
\end{equation}
can be approximated to linear order in $|\lambda|\lVert B\otimes A_i\rVert$. Here and in the following we use the symbol $\simeq$ to denote validity up to linear order in $\lambda$. Physically, the required condition $|\lambda|\lVert B\otimes A_i\rVert\ll1$ can be satisfied either by implementing a Hamiltonian of weak interaction strength $\lVert B\otimes A_i\rVert$, and/or by choosing the coupling time $\lambda$ sufficiently small. Here we will take the point of view that $|\lambda|\ll1$ and choose, without loss of generality, coupling operators such that $\lVert A_i\rVert=1$ and $\lVert B\rVert=1$. At the end of the coupling procedure, one obtains
\begin{equation}\label{eq: post_weak_coupling_state}
\ket{\Psi_\lambda(t_1)} \simeq \ket{\phi,\psi(t_1)} - i \lambda \ket{B \phi, A_i \psi(t_1)}.
\end{equation}

\paragraph{Measuring the ancilla.}
The state of the ancilla is then probed by projectively measuring the observable $S^a\otimes\id_{\text{S}}$, i.e., for the ancilla spin, the same component $a$ that occurs in the correlation function \eqref{e:C_th} at lattice site $i$ is probed. We denote the $2s+1$ eigenstates of $S^a$ as $\ket{m_a}$ with corresponding eigenvalues $m_a \in \mathscr{S}$. According to the Born rule, one measures $m_a$ with probability
\begin{multline}\label{eq: NIM_probt1}
	P_{m_a} \simeq \matel{\Psi_\lambda(t_1)}{\left( \projector{m_a} \otimes \id_{\text{S}} \right)}{\Psi_\lambda(t_1)}\\
	= \left\vert \innprod{m_a}{\phi} \right\vert^2 - i\lambda \expval{A_i(t_1)}_{\psi} \left( \innprod{\phi}{m_a} \matel{m_a}{B}{\phi}-\text{c.c.} \right),
\end{multline}
where $\text{c.c.}$ denotes the complex conjugate and $\expval{A_i(t_1)}_{\psi}=\matel{\psi}{U^\dagger(t_1) A_i U(t_1)}{\psi}$.
The post-measurement state is given by the normalized (and linearized with respect to $\lambda$) projection onto the subspace corresponding to the outcome $m_a$ of the measurement,
\begin{equation}
\begin{split}
\ket{\Psi_{m_a}(t_1)} &\simeq \frac{\left( \ket{m_a}\bra{m_a} \otimes \id_{\text{S}} \right)\ket{\Psi_\lambda(t_1)}}{\left\lVert\left( \ket{m_a}\bra{m_a} \otimes \id_{\text{S}} \right)\ket{\Psi_\lambda(t_1)}\right\rVert}\\ 
&\simeq \ket{m_a}\otimes \ket{\psi_{m_a}(t_1)}
\end{split}
\end{equation}
with
\begin{multline}\label{e:postMeasLin}
	\!\!\ket{\psi_{m_a}(t_1)} \simeq \Biggl\{\dfrac{ \innprod{m_a}{\phi} }{ \left\vert \innprod{m_a}{\phi} \right\vert } - i \lambda \Biggl[ \dfrac{ \matel{m_a}{B}{\phi} }{ \left\vert \innprod{m_a}{\phi} \right\vert } A_i - \dfrac{ \innprod{m_a}{\phi} }{ 2\left\vert \innprod{m_a}{\phi} \right\vert^3 }\\
	\times\expval{A_i(t_1)}_{\psi} \left( \innprod{\phi}{m_a} \matel{m_a}{B}{\phi} - \text{ c.c.} \right) \Biggr]\Biggr\} \ket{\psi(t_1)}.
\end{multline}
Ancilla and system are again in a product state.

\paragraph{Time evolution until time $t_2$.}
Time-evolve the post-measurement state $\ket{\Psi_{m_a}(t_1)}$ up to the time $t_2$ with the system Hamiltonian $H_S$,
\begin{equation}
\ket{\Psi_{m_a}(t_2)} \simeq \ket{m_a}\otimes e^{-i H_s (t_2-t_1)}\ket{\psi_{m_a}(t_1)}.
\end{equation}
The ancilla state $\ket{m_a}$ remains unaffected.

\paragraph{Projective measurement at site $j$.}
At the final time $t_2$, the disturbing effect due to a measurement is not of concern, and we can projectively measure the observable $S_j^b$ at lattice site $j$ without compromising the accuracy of the correlation function \eqref{e:C_th} which we wish to measure. The conditional probability of measuring the system in eigenstate $\ket{m_b}$ of $S_j^b$ after having obtained eigenvalue $m_a$ when measuring the ancilla is
\begin{align}\label{eq: NIM_conditional_probt2}
	P_{m_b \vert m_a} &\simeq \matel{\Psi_{m_a}(t_2)} {(\mathds{1}_{\text{A}} \otimes \ket{m_b} \bra{m_b})} {\Psi_{m_a}(t_2)}\nonumber\\
	&\simeq \left\vert \matel{m_b}{U(t_2)}{\psi} \right\vert^2 - i \lambda \Biggl\lbrace \dfrac{\innprod{\phi}{m_a} \matel{m_a}{\hat{B}}{\phi} }{ \left\vert \innprod{m_a}{\phi} \right\vert^2 }\nonumber\\
	&\quad \times \Bigl[\matel{\psi}{U^\dagger(t_2)}{m_b} \matel{m_b}{ U(t_2-t_1) A_i U(t_1)}{\psi}\nonumber\\
	&\quad- \expval{ A_i(t_1) }_{\psi} \left\vert \matel{m_b}{ U(t_2)}{\psi} \right\vert^2
	\Bigr] - \text{c.c.} \Biggr\rbrace.
\end{align}

\paragraph{Correlating the measured outcomes.}
\label{p:g}
We use the probabilities \eqref{eq: NIM_probt1} and \eqref{eq: NIM_conditional_probt2} to calculate the correlation \eqref{e:C_proj} between the measured ancilla spin at $t_1$ and the system spin $j$ at $t_2$,
\begin{equation}\label{eq: wmcf}
\begin{split}
\wmcf(t_1,t_2) =& \sum_{ m_a, m_b \in \mathscr{S} } m_a m_b P_{m_b \vert m_a} P_{m_a} \\
\simeq& \expval{S^a}_{\phi} \expval{S_j^b(t_2)}_{\psi}\\
&- i \lambda \left[ \expval{S^a B}_{\phi}\matel{\psi}{ S_j^b(t_2) A_i(t_1) }{\psi} - \text{c.c.}  \right],
\end{split}	
\end{equation}
where we have absorbed the summations via the spectral representations of $S^a$ and $S_j^b$.
By setting $A_i=S_i^a$, the last line in \eqref{eq: wmcf} is made to contain the desired correlation \eqref{e:C_th}, which, in light of the fact that the ancilla has been measured projectively, is a remarkable finding. 

Isolating this desired term requires exact knowledge of the value $\expval{S^a}_{\phi} \expval{S_j^b(t_2)}_{\psi}$. Since the initial system state $\ket{\psi}$ is generally unknown, the best strategy is to choose the initial ancilla state
\begin{equation}\label{e:phi}
\ket{\phi} = \sum_{m_a \in \mathscr{S}} c_{m_a} \ket{m_a}
\end{equation} 
such that 
\begin{equation}\label{e:expCoeff}
\expval{S^a}_{\phi}=0,
\end{equation}
which is satisfied if the coefficients $c_{m_a}$ in \eqref{e:phi} satisfy
\begin{equation}
\sum_{m_a \in \mathscr{S}, m_a > 0} m_a \left( \vert c_{m_a} \vert^2 - \vert c_{-m_a} \vert^2 \right) = 0.
\end{equation}
Physically relevant states satisfying this condition are, for instance, spin coherent states, or equal superpositions where $c_{m_a} = 1/\sqrt{2s+1}$ for all $m_a \in \mathscr{S}$. We choose the latter for our derivation, noting that other choices only lead to modified prefactors $f^{(1)},f^{(2)}$ in \eqref{e:CB1} and \eqref{e:CB2}. 

With condition \eqref{e:expCoeff} satisfied, \eqref{eq: wmcf} reduces to
\begin{equation}\label{e:16}
	\wmcf(t_1,t_2) \simeq \frac{-2 \lambda}{2s+1}\!\sum_{m_a,m'_a \in \mathscr{S}}\! m_a \impart{\matel{m_a}{B}{m'_a}C(t_1,t_2)},
\end{equation}
from which we can extract the real or imaginary part of $C$ through suitable choices of $B$. Choosing $B$ Hermitian and symmetric renders \eqref{e:16} proportional to the imaginary part of $C$. A physically natural choice is
\begin{equation}\label{e:B1}
B=B^{(1)}=S^a,
\end{equation}
which yields
\begin{equation}\label{e:CB1}
	\wmcf^{(1)}(t_1,t_2) \simeq -\frac{2\lambda f^{(1)}}{2s+1} \impart{C(t_1,t_2)}
\end{equation}
with $f^{(1)}=\sum_{m_a \in \mathscr{S}} m_a^2$. Choosing $B$ Hermitian and antisymmetric makes \eqref{e:16} proportional to the real part of $C$. For $S^a=S^z$ a physically appealing choice is $B=S^y$ or, analogously for general $a\in \lbrace x,y,z \rbrace$, the spin component
\begin{equation}\label{e:B2}
B = B^{(2)} = -\frac{i}{2}\left(S_a^+-S_a^-\right),
\end{equation}
where $S_a^\pm$ denote spin-lowering or -raising operators with respect to the $m_a$-eigenbasis.
Then \eqref{e:16} reduces to
\begin{equation}\label{e:CB2}
	\wmcf^{(2)}(t_1,t_2) \simeq -\frac{2\lambda f^{(2)}}{2s+1} \realpart{C(t_1,t_2)}
\end{equation}
with $f^{(2)}=i\sum_{m_a,m'_a \in \mathscr{S}}m_a\matel{m_a}{B^{(2)}}{m'_a}$.
Inverting Eqs.~\eqref{e:CB1} and \eqref{e:CB2}, we can define
\begin{equation}\label{e:C}
C^\lambda(t_1,t_2) = -\frac{2s+1}{2\lambda}\left(\frac{\wmcf^{(2)}(t_1,t_2)}{f^{(2)}}+i\frac{\wmcf^{(1)}(t_1,t_2)}{f^{(1)}} \right),
\end{equation}
which approximates the exact correlation function $C(t_1,t_2)$ for sufficiently small $\lambda$.

Equation \eqref{e:C} is the first main result of this paper, demonstrating the validity of the proposed noninvasive measurement protocol. 
It shows that experimental implementation, discussed further in Sec.~\ref{sec:exp_imp}, will require two measurement samples, one for system--ancilla coupling $B^{(1)} \otimes S^a_i$ and a second one for $B^{(2)} \otimes S^a_i$, in order to construct the complex-valued correlation function \eqref{e:C}.
It is remarkable that the first-order (in $\lambda$) approximation of the ancilla--system coupling $\couple$ leads to such a succinct relation between $C(t_1,t_2)$ and $\wmcf(t_1,t_2)$. The protocol can be applied to any spin model regardless of interaction type, spin number or dimensionality. 

A number of measurement schemes discussed in the literature bear some superficial similarity to the above described protocol. In Ref.~\cite{JohansenMello08} two noninvasive measurements are made in succession, but not in a way suitable for, nor with the aim of, allowing for the full reconstruction of dynamical correlation functions. Other references use noninvasive measurements to show violations of Leggett-Garg inequalities, but the latter are inequalities for the dynamic correlations of (real) measurement outputs, so connecting the result to the (complex) dynamic correlation function \eqref{e:C_th} is not part of the agenda \cite{JordanKorotkovBuettiker06,Goggin_etal11}. 

\section{Finite-sample estimators and errors}
\label{sec:estimator_construction}

The key formula \eqref{e:C} of the noninvasive measurement protocol contains 
the system--ancilla dynamical correlation function $\wmcf$ defined in \eqref{eq: wmcf}, which in turn requires the knowledge of the outcome probabilities $P_{m_a}$ and $P_{m_b \vert m_a}$. An exact calculation of these probabilities, which involve the time-evolution under the many-body Hamiltonian $H$, is in almost all cases impossible. Experimentally one can estimate the probabilities by doing multiple repetitions of the protocol of Sec.~\ref{sec:NIM_protocol}, and then combine the estimated probabilities according to \eqref{eq: wmcf} to obtain estimators $\wmcf^{(m)}_n$ of the system--ancilla correlation function $\wmcf^{(m)}$ with $m=1,2$,
where the subscript $n$ indicates the use of a finite sample of $n$ measurements.

Due to the finite sample size, the estimators will be error-prone, and this error propagates into the estimated dynamic correlation function
\begin{equation}\label{e:Cn}
C^\lambda_n(t_1,t_2) = -\frac{2s+1}{2\lambda}\left(\frac{\wmcf^{(2)}_n(t_1,t_2)}{f^{(2)}}+i\frac{\wmcf^{(1)}_n(t_1,t_2)}{f^{(1)}} \right).
\end{equation}
From Eq.~\eqref{e:Cn} it follows that the noise contained in the signal $\wmcf_n$ will be inherited by $C^\lambda_n$ and, by standard error propagation, will be strongly amplified in the limit $\lambda\to0$.

At this point an interesting optimization problem arises: The noninvasive measurement protocol of Sec.~\ref{sec:NIM_protocol} was derived in linear order in $\lambda$, and is hence accurate only for sufficiently weak system--ancilla couplings $\lambda$, while larger $\lambda$ will lead to {\em systematic}\/ errors in the estimators $\wmcf^{(m)}_n$, and hence in $C^\lambda_n$. The {\em statistical}\/ errors discussed in the previous paragraph show the opposite tendency, becoming smaller with increasing $\lambda$. The {\em total}\/ error in $C^\lambda_n$, given by the sum of systematic and statistical errors, is therefore expected to take on a minimum at some intermediate value $\lambda^*$ of the ancilla--system coupling. Since the systematic error is independent of the sample size $n$, while the statistical error decreases with increasing $n$, we expect $\lambda^*$ to decrease as $n$ increases. Realistically, however, limited resources (man power or time or money), will cap the maximum sample size $n$.

For the application of the noninvasive measurement protocol, the following optimization problem is therefore of relevance: Given a finite sample size $n$, what is the optimal $\lambda$ such that the sum of systematic and statistical error becomes minimal? 

In the remainder of this section we investigate this question by deriving a bound on the total error. For doing so, it may be convenient to recapitulate the different (estimators of) correlation functions that we have introduced.
\renewcommand{\descriptionlabel}[1]{\hspace{\labelsep}\emph{#1}:}
\begin{description}
\item[$C$] Exact correlation function \eqref{e:C_th}; this is the quantity we would like to extract by means of noninvasive measurements.
\item[$C^\lambda$] Correlation function \eqref{e:C}, defined in terms of the probabilities of system and ancilla measurement outcomes as in the second line of \eqref{eq: wmcf}. Shown to be equal to $C$ asymptotically in the limit of small $\lambda$. In principle, an infinite number of measurements would be required to determine the exact probabilities.
\item[$C^\lambda_n$] Correlation function \eqref{e:Cn}, defined like $C^\lambda$ in terms of system and ancilla measurement outcomes, but with probabilities replaced by relative frequencies. This is the quantity one actually obtains from a sequence of $2n$ measurements ($n$ for each operator $B^{(m)}$).
\end{description}
The systematic, statistical, and total errors are then respectively given by 
\begin{subequations}
\begin{align}
	\eSyst &= \left|C-C^\lambda\right| \label{eq: error_syst},\\
	\eFS &=  \left|C^\lambda-C^\lambda_n\right| \label{eq: error_finite_size},\\
	\estDev &= \left|C-C^\lambda_n\right| \leq \left|\eSyst\right|+\left|\eFS\right|. \label{eq: total_deviation}
\end{align}
\end{subequations}

The statistical error originates from the replacement of probabilities in the first line of \eqref{eq: wmcf} by the corresponding relative frequencies with which the different outcomes are measured in a sequence of $n$ measurements. This replacement is most directly done in the non-conditional probabilities
\begin{equation}
P_{m_a m_b}=P_{m_a}P_{m_b|m_a},
\end{equation}
which denote the joint probabilities of measuring $m_a$ for the ancilla spin and $m_b$ for the system spin at site $j$. In a sample of $n$ measurements, one will observe the $(2s+1)^2$ possible outcome combinations $(m_a,m_b)$ with relative frequencies $n_{m_a m_b}/n$, such that
\begin{equation}
\sum_{m_a,m_b} n_{m_a m_b}=n\quad\text{and}\quad\lim_{n\rightarrow\infty} \dfrac{n_{m_a m_b}}{n} = P_{m_a m_b}.
\end{equation}
For sufficiently large $n$, one expects $n_{m_a m_b}/n$ to be Poisson-distributed with mean $P_{m_a m_b}$ and standard deviation $\sqrt{n_{m_a m_b}}/n$ \cite{HDYoung}. Making use of $n_{m_a m_b}/n=P_{m_a m_b}\pm\sqrt{n_{m_a m_b}}/n$, we find
\begin{equation}\label{e:qaz}
\wmcf_n = \sum_{m_a,m_b}\! m_a m_b \frac{n_{m_a m_b}}{n} = \wmcf + \sum_{m_a,m_b}\! m_a m_b \frac{\pm \sqrt{n_{m_a m_b}}}{n}.
\end{equation}
Substituting \eqref{e:Cn} and \eqref{e:C} into \eqref{eq: error_finite_size} we find
\begin{equation}\label{e:estat_bound}
\begin{split}
\eFS &\leq \frac{2s+1}{2 \vert \lambda \vert} \left( \frac{\vert \wmcf^{(2)} - \wmcf^{(2)}_n \vert}{f^{(2)}} + \frac{\vert \wmcf^{(1)} - \wmcf^{(1)}_n \vert}{f^{(1)}} \right)\\
&\leq \frac{2s+1}{2 \vert \lambda \vert} \sum_{m_a,m_b} \vert m_a m_b\vert \left( \frac{ \sqrt{ n^{(2)}_{m_a m_b}} }{f^{(2)}} + \frac{\sqrt{ n^{(1)}_{m_a m_b}}}{f^{(1)}} \right),
\end{split}
\end{equation}
where \eqref{e:qaz} and the triangle inequality were used. From this estimate we expect that, for a fixed sample size $n$, the noise-to-signal ratio of the noninvasive measurement protocol diverges in the limit of small $\lambda$.

Estimating the systematic error $\eSyst$ is much more challenging in general, as it involves the exact dynamic correlation function $C$, which is usually unknown. One possible approach is to redo the calculations of Sec.~\ref{sec:NIM_protocol} to next-to-leading order in $\lambda$, from which we could estimate the linear (in $\lambda$) contribution to $\eSyst$ in the regime of small $\lambda$. In the next section we will follow a different approach, trying to obtain an understanding of the interplay between systematic and statistical errors by discussing an exactly solvable minimal model, consisting of three spin-$1/2$ particles: one ancilla and two system degrees of freedom.

\section{Example: two system spins, one ancilla}
\label{sec:numeric_imp}

As a minimal model for investigating spatio-temporal correlations by means of noninvasive measurements, we require a system consisting of two sites, plus a single ancilla spin. The resulting Hilbert space of three spin-$1/2$ degrees of freedom is eight-dimensional, and all calculations can be performed numerically with little effort.

We choose a Hamiltonian with Ising-type spin--spin coupling,
\begin{equation}\label{e:Hxx}
H=\sigma_1^x \sigma_2^x,
\end{equation}
and consider dynamics starting from an initial product state
\begin{equation}
\ket{\Psi} = \ket{\phi}\otimes\ket{\psi_1}\otimes\ket{\psi_2},
\end{equation}
where the system spin states are parametrized by angles $\alpha_i \in [0,\pi/2]$ and $\theta_i\in [0,2\pi]$ as
\begin{equation}\label{eq: initial_system_state}
\ket{\psi_i} = \cos(\alpha_i) e^{-i \theta_i/2} \ket{+_z} + \sin(\alpha_i) e^{i \theta_i/2} \ket{-_z}
\end{equation}
for $i=1,2$, and the ancilla initial state $\ket{\phi}$ is given by \eqref{e:phi}.
Our aim is to apply the noninvasive measurement protocol for estimating the dynamical $zz$ correlation function
\begin{multline}\label{eq:C_ex}
C(t_1,t_2) = \matel{\psi_1\psi_2}{\sigma_1^z (t_1) \sigma_2^z (t_2)}{\psi_1\psi_2}\\
=\cos(2 \alpha_1) \cos(2 \alpha_2) \cos(2 (t_2-t_1)) \\
+ i \sin(2 \alpha_1) \sin(2 \alpha_2) \sin(\theta_1) \sin(\theta_2) \sin(2 (t_2-t_1)) .
\end{multline}

\begin{figure*}\centering
  \includegraphics[width=0.33\linewidth]{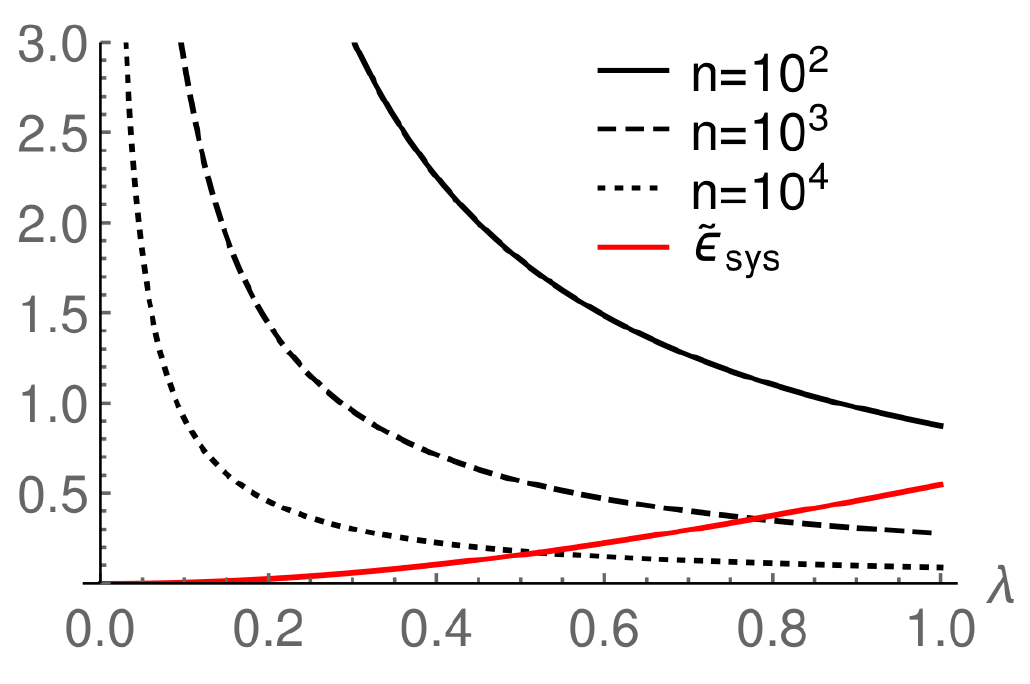}
  \hfill
  \includegraphics[width=0.33\linewidth]{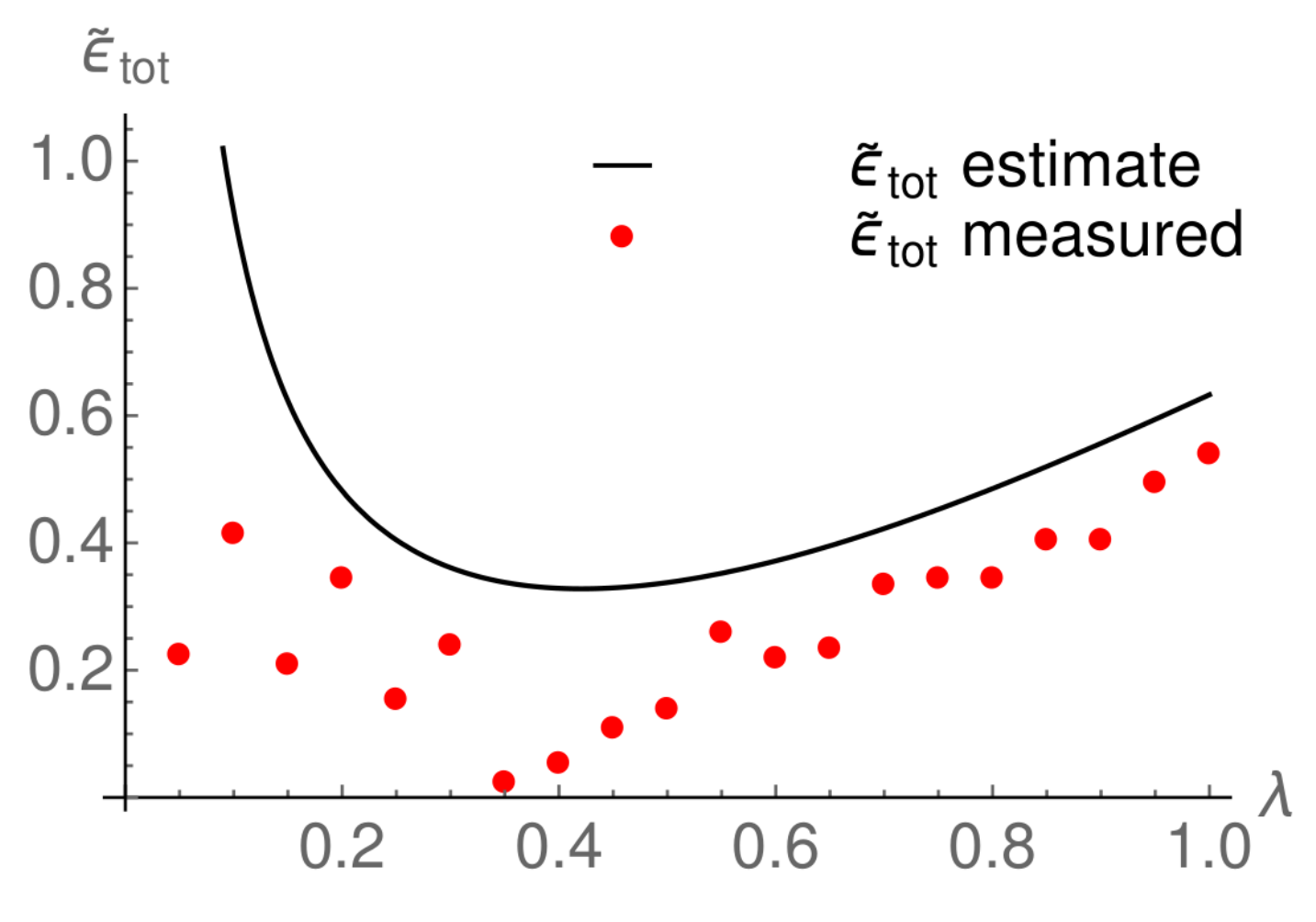}
  \hfill
  \includegraphics[width=0.33\linewidth]{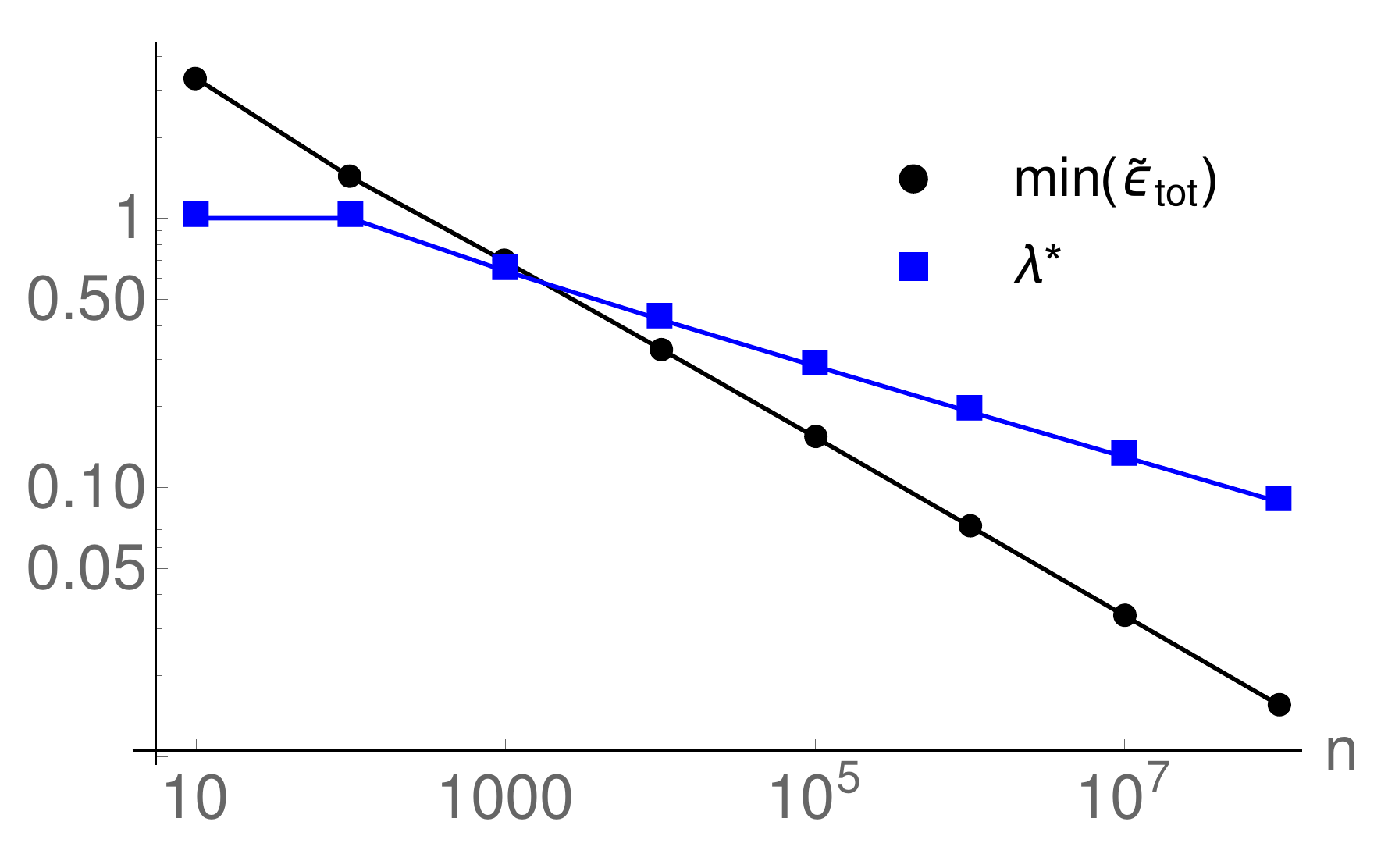}
\caption{\label{fig: predicted_errors}%
Error analysis for the example of Sec.~\ref{sec:numeric_imp} for parameter values $(t_1,t_2)=(1,10)$, $\alpha_1=\alpha_2=\pi/3$ and $(\theta_1,\theta_2)=(\pi/7,\pi/5)$. All plots show relative errors $\tilde{\epsilon}_\text{sys}=\epsilon_\text{sys}/|C|$, and similarly for statistical and total errors.
Left: Exact systematic error $\tilde{\epsilon}_\text{sys}$ (red curve starting from the origin) and estimates \eqref{e:estat_bound} of $\tilde{\epsilon}_\text{stat}$ for sample sizes $n=10^2,10^3,10^4$ (solid, dashed, dotted black curves respectively). The functional dependence of the bounds on the coupling time $\lambda$ is clear, with the systematic error increasing with $\lambda$, while the stochastic error is inversely proportional to $\lambda$.
Center: Estimate for $\tilde{\epsilon}_\text{tot}$ (line) together with numerically measured values for sample size $n=10^4$ (dots). The estimated minimum error is $33\% $ at $\lambda^*=0.42$. The error estimate captures the qualitative behavior of the numerical data and is consistently larger.
\label{fig: predErr_manySamples}
Right: Log-log plot of the estimated minimum error $\tilde{\epsilon}_\text{tot}(\lambda^*)$ (dots) and its position $\lambda^*$ (squares) as a function of sample size $n$. Straight lines indicate that the performance of the noninvasive measurement protocol improves like a power law with $n$.}
\end{figure*}

To obtain the systematic error $\epsilon_{\text{sys}}$ \eqref{eq: error_syst} one needs the probabilities occurring in \eqref{eq: wmcf} as they arise in the protocol of Sec.~\ref{sec:NIM_protocol}, but without linear approximations in $\lambda$. Calculating these probabilities for the Hamiltonian \eqref{e:Hxx} and combining them according to the middle line of \eqref{eq: wmcf} we can to all orders in $\lambda$ construct
\begin{equation}\label{e:Clambda_ex}
\begin{split}
\!\!\!\!\!\!\!\!C^\lambda(t_1,t_2) = \dfrac{1}{2 \lambda}\!\Bigl(\!\cos(2 \alpha_1) \cos(2\alpha_2) \sin(2\lambda)\cos(2(t_2-t_1))\\ 
+ i \sin(2\alpha_1) \sin(2\alpha_2) \sin(\theta_1)\sin(\theta_2) \sin(2\lambda) \sin(2(t_2-t_1)) \!\Bigr)
\end{split}
\end{equation}
as defined in \eqref{e:C}. Substituting \eqref{eq:C_ex} and \eqref{e:Clambda_ex} into \eqref{eq: error_syst} we obtain the systematic error
\begin{multline}\label{eq: syst_error_calc}
\!\!\!\!\!\!\!\eSyst = \dfrac{1}{2 \vert\lambda\vert} \Bigl\vert (2\lambda-\sin(2 \lambda)) \Bigl[ \cos(2(t_2-t_1))\cos(2\alpha_1)\cos(2\alpha_2) \\
+i  \sin(2(t_2-t_1)) \sin(\theta_1) \sin(\theta_2)\sin(2\alpha_1)\sin(2\alpha_2) \Bigr] \Bigr\vert.
\end{multline}
 The systematic error $\epsilon_{\text{sys}}$, which vanishes for $\lambda \rightarrow 0$, is shown in Fig.~\ref{fig: predicted_errors} (left, red line increasing from origin) for the example $\alpha_1 = \alpha_2 = \pi/3$ and $\lambda \in (0,1]$. In the same plot the upper bound \eqref{e:estat_bound} on the statistical error, which decreases with increasing $\lambda$, is shown for various sample sizes $n$ (black lines as indicated by the legend). For sufficiently large $n$ the total error bound $\epsilon_\text{tot}$ shows a minimum for some $\lambda=\lambda^*$ (Fig.~\ref{fig: predicted_errors}, center). Hence, assuming the bounds to be reasonably tight, $\lambda=\lambda^*$ should be a good choice for the system--ancilla coupling when using  a sample of $n= 10^4$ measurements. The corresponding error estimate is fairly large 
due to the conservative upper bound of the statistical error \eqref{e:estat_bound}.

To check the tightness of the bounds, we numerically implemented the noninvasive measurement protocol of Sec.~\ref{sec:NIM_protocol}, using the exact time evolution (without expanding in $\lambda$) and drawing samples of random numbers according to the ancilla- and system-spin outcome probabilities \footnote{The relative frequencies $n_i/n$ needed to determine \eqref{e:Cn} were obtained by binning $2 \times 10^4$ pseudo-random numbers  drawn from the unit interval, with the probabilities $P_i$ determining the bin widths.}. As expected, the results of the numerical implementation (red dots, Fig.~\ref{fig: predicted_errors} center) are smaller than the conservatively estimated analytical error bounds. The influence of the statistical and systematic errors is evident in the numeric data. For $\lambda < \lambda^*$ the statistical error dominates, causing fluctuations whose sizes are of the same order as the total error. For $\lambda > \lambda^*$ the measured errors exhibit smaller fluctuations, and follow the trend of the error bound, reflecting the increasingly dominant role of the systematic error at larger $\lambda$.
To consistently achieve good accuracies, the ancilla-system coupling at $t_1$ should be chosen close to, but not smaller than $\lambda^*$. 

The performance of the noninvasive protocol is characterized in Fig.~\ref{fig: predicted_errors} (right), where the minimum value $\tilde{\epsilon}_\text{tot}(\lambda^*)$ as well as the corresponding $\lambda^*$ is shown for a range of sample sizes. The minimum error decays like a power law with increasing sample size, and so does the corresponding optimal coupling $\lambda^*$. This plot answers, at least on the level of error estimates, the optimization question posed at the beginning of Sec.~\ref{sec:estimator_construction}. The conservative error estimates assume individual errors to not compensate each other, and experimental implementations are therefore expected to achieve smaller errors.

\section{Generalisations}
\label{s:generalizations}

The noninvasive measurement protocol of Sec.~\ref{sec:NIM_protocol} is based on two key ingredients: weak system--ancilla coupling to reduce measurement backaction, and multiple repetitions of the protocol to achieve a sufficient signal-to-noise ratio. To further improve the protocol, one may wonder whether measurement backaction can be further reduced by coupling (and then decoupling) an ancilla to the system at time $t_1$, but measuring the ancilla (step $d.$ in the protocol of Sec.~\ref{sec:NIM_protocol}) only at time $t_2$ or even later. Such deferred measurements, recently suggested in \cite{Oehri_etal16}, are shown in Appendix \ref{s:deferred} to have no effect on our suggested protocol and lead to the exact same estimators and errors.

Another modification of the protocol could seek to make better use of experimental resources by performing multiple noninvasive measurements at times $t_1$, $t_2$, $t_3$, \dots with the aim of extracting several dynamic correlation functions $C(t_1,t_2)$, $C(t_1,t_3)$, $C(t_2,t_3)$, \dots from the same sample of experimental runs. We show in Appendix \ref{sec:multiple_times} that such a protocol is feasible, but it turns out to be less efficient than separate noninvasive measurements for the desired dynamic correlation functions.

\section{Implementation in linear ion traps}
\label{sec:exp_imp}

The noninvasive measurement protocol of Sec.~\ref{sec:NIM_protocol} requires a high level of control, in particular the possibility to prepare an ancilla in a well-defined state and to couple it to and decouple it from the system. There are a number of experimental platforms based on trapped cold atoms, molecules, or ions with which quantum spin models can be emulated, and which hold the potential for implementing the protocol. Here we discuss in some detail a scheme suitable for trapped ions in linear radio-frequency traps (Paul traps) and show that all required steps can be implemented with current technology.

Choosing two hyperfine electronic states of a trapped ion as spin states $\ket{\pm}$, one can drive transitions from one to the other with an oscillating field whose frequency is tuned to the energy gap of the two states \cite{Ozeri2011}. As a result each ion can be modeled as an effective spin-$1/2$ particle.
Furthermore, linear Paul traps confine ions along a single axis in real-space \cite{Steane1997}, thereby creating a one-dimensional array of $N$ ions, which can be modeled as a chain of $N$ spin-1/2 particles. 

We propose to designate one of the $N$ trapped ions as the spin-$1/2$ ancilla particle, which can be prepared in the required initial state \eqref{e:phi} via single-ion laser addressing provided that the qubit transition is in the optical regime. The remaining ions form the system, and their dynamics under action of a desired Hamiltonian can be initiated and driven for some time $t_1$. To measure dynamic correlations between the ions at sites $i$ and $j$, it is not necessary for the ancilla ion to be adjacent to any of these sites; it may be located at any site, including the chain ends. To retain the initial ancilla state, one must exclude it from the dynamics. This can be achieved with a ``spin-shelving'' procedure \cite{Steane1997}, which involves placing the ancilla ion in a different external state such that the dynamics-generating driving field does not couple to that ion.

To realize step $c.$\ of the protocol in Sec.~\ref{sec:NIM_protocol}, the system dynamics must be temporarily stopped at time $t_1$ by switching off the driving fields, and the ancilla spin must be coupled with only the $i$th lattice spin. Interactions between ions are mediated through collective phonon modes of the ion lattice. Restricting the interaction to a specific ion pair requires sophisticated but well-established techniques, as described in \cite{MolSor, Leibfried,Sawyer_etal,Kim_etal09}.
One suitable technique to generate the system--ancilla coupling \eqref{eq: coupling_linear} under Hamiltonian \mbox{$H_c = B \otimes A_i$} is a method in which entangling gates are mediated by phonon modes transverse to the trap axis \cite{Kim_etal09,Zhu_etal}. For a linear ion trap, these transverse phonon mode gates can entangle the spin states of arbitrary ion pairs and for a chosen coupling strength $\lambda$, by producing a $\sigma^z\sigma^z$ interaction.

The system--ancilla coupling is then restricted to $\mathscr{U}(\lambda) = \exp(-i\lambda \sigma^z \otimes \sigma_i^z)$, which is sufficient for measuring the imaginary part of $\matel{\psi}{\sigma_i^z(t_1) \sigma_j^b(t_2)}{\psi}$ with $b \in \lbrace x,y,z  \rbrace$, but not for other spin components.
We show in Appendix~\ref{app:ancillaRot} that appropriate rotations of the system and the ancilla allow us to extract estimators for real and imaginary parts of arbitrary dynamic correlations $\matel{\psi}{\sigma_i^a(t_1) \sigma_j^b(t_2)}{\psi}$.

\section{Projective measurement protocol for spin-\texorpdfstring{$1/2$}{1/2} models}
\label{s:DPMP}
The noninvasive measurement protocol of Sec.~\ref{sec:NIM_protocol}, which is valid for general spin models and observables, was introduced with the aim of reducing, and essentially eliminating in the limit of small $\lambda$, the disturbing effect of measurement backaction at the early measurement time $t_1$. 
Surprisingly, for spin-$1/2$ models, but with otherwise general Hamiltonians, dynamic correlations
\begin{align}\label{e:C_th_spinhalf}
C(t_1,t_2) &= \left\langle \sigma_i^a(t_1)\sigma_j^b(t_2)\right\rangle
\end{align}
can be obtained with strictly zero disturbance from measurement backaction. This can be achieved for the real part of $C$ by the following protocol, based on projective measurements, and for the imaginary part by a protocol based on local rotations, as put forward in Sec.~\ref{s:RPMP}.

\paragraph{Time evolution until time $t_1$,}
\begin{equation}
\ket{\psi(t_1)}=e^{-iH t_1}\ket{\psi}.
\end{equation}

\paragraph{Projective measurement at site $i$.}
The state is probed by projectively measuring the observable $\sigma_i^a$. We denote the eigenstates of $\sigma_i^a$ as $\ket{+_a}$ and $\ket{-_a}$ with eigenvalues $+1$ and $-1$, respectively, and the corresponding projectors by $\Pi_i^{\pm_a}$. According to the Born rule, one measures $\pm1$ with probabilities
\begin{align}\label{eq:dp_p1}
P_{\pm_a}^\text{proj} = \bra{\psi(t_1)} \Pi_i^{\pm_a} \ket{\psi(t_1)},
\end{align}
and the corresponding post-measurement states are
\begin{align}
\ket{\psi_{\pm_a} (t_1)} = \dfrac{\Pi_i^{\pm_a}\ket{\psi(t_1)}}{\Vert \Pi_i^{\pm_a} \ket{\psi(t_1)} \Vert} = \dfrac{\Pi_i^{\pm_a}\ket{\psi(t_1)}}{\sqrt{P_{\pm_a}^\text{proj}(t_1)}}.
\end{align}

\paragraph{Time evolution until time $t_2$.}
Time-evolve the post-measurement state $\ket{\psi_{\pm_a}(t_1)}$ up to the time $t_2$,
\begin{equation}
\ket{\psi_{\pm_a}(t_2)} = e^{-iH(t_2-t_1)}\ket{\psi_{\pm_a}(t_1)}.
\end{equation}

\paragraph{Projective measurement at site $j$.}
The conditional probability of measuring the system in eigenstate $\ket{\pm_b}$ of $\sigma_j^b$ at time $t_2$ after having obtained $\ket{\pm_a}$ when measuring $\sigma_i^a$ at time $t_1$ is
\begin{align}\label{eq:dp_p2}
P_{\pm_b \vert \pm_a}^\text{proj} = \matel{\psi_{\pm_a} (t_2)}{\Pi_j^{\pm_b}}{\psi_{\pm_a} (t_2)}.
\end{align}

\paragraph{Correlating the measured outcomes.}
Correlations are calculated according to 
\begin{equation}\label{e:ps-pd}
\mathscr{C}^\text{proj} = P_{+_a +_b}^\text{proj}+P_{-_a -_b}^\text{proj}-P_{-_a +_b}^\text{proj}-P_{+_a -_b}^\text{proj},
\end{equation}
where $P_{+_a +_b}^\text{proj}=P_{+_a}^\text{proj}P_{+_b|+_a}^\text{proj}$ denotes the joint probability to projectively measure outcome $+$ for $\sigma_i^a$ at time $t_1$ and outcome $+$ for $\sigma_j^b$ at $t_2$ (and similarly for the other indices). Inserting \eqref{eq:dp_p1} and \eqref{eq:dp_p2} into \eqref{e:ps-pd} and after some algebraic manipulations (reported in Appendix \ref{app:naive_general}) we obtain the final result 
\begin{multline}\label{e:c_dpVSth}
\wmcf^\text{proj}(t_1,t_2) = C(t_1,t_2)\\
+ 2 i \impart{ \matel{\psi}{ \Pi_i^{-_a} (t_1) \sigma_j^b(t_2) \Pi_i^{+_a} (t_1)}{\psi} },
\end{multline}
where we have abbreviated $U^\dagger(t_1) \Pi_i^{\pm_a} U(t_1)= \Pi_i^{\pm_a} (t_1)$.
$\wmcf^\text{proj}$ is real per its definition \eqref{e:ps-pd} and the second term on the right-hand side of \eqref{e:c_dpVSth} is purely imaginary. Hence it follows that 
\begin{equation}\label{e:dp_real}
\wmcf^\text{proj}(t_1,t_2) = \realpart{C(t_1,t_2)}.
\end{equation}

For Hamiltonians beyond spin-$1/2$ and/or for general observables, such a projective measurement protocol does not yield the desired dynamic correlation functions. More precisely, we have shown that $\realpart{ C}=\mathscr{C}^\text{proj}$ holds only if the operator
\begin{equation}\label{e:projCond}
\Gamma \equiv \sum_{m_a} m_a \sum_{m'_a \neq m_a} \Pi_i^{m_a} (t_1) S_j^b(t_2) \Pi^{m'_a} (t_1)
\end{equation}
is anti-hermitian; see Appendix~\ref{app:naive_general} for a proof. This condition is satisfied for the spin-$1/2$ setting considered above, but is violated in most other cases, for example for spin-$1$ models (see Appendix~\ref{app:naive_general}).

\section{Rotation-based measurement protocol for spin-\texorpdfstring{$1/2$}{1/2} models}
\label{s:RPMP}
For spin-$1/2$ models, the imaginary part of $\matel{\psi}{\sigma^a_i(t_1)\sigma_j^b(t_2)}{\psi}$ can likewise be obtained without the use of ancillas and with strictly zero effect from measurement backaction. This is achieved by the following measurement protocol, based on local rotations.

\paragraph{Time evolution until time $t_1$,}
\begin{equation}
\ket{\psi(t_1)}=e^{-iH t_1}\ket{\psi} .
\end{equation}

\paragraph{Local rotation at site $i$.}
The $i^{\text{th}}$ lattice-spin is rotated, parallel to the axes of the spin component which is to be correlated at $t_1$, by applying the unitary
\begin{equation}
R_i(\theta,\mvec{a}) = e^{-i \frac{\theta}{2} \sigma_i^a} = \cos(\theta/2) - i\sin(\theta/2) \sigma_i^a .
\end{equation}
The locally rotated system state is then
\begin{equation}
\ket{\psi_{\theta}(t_1)} = \left(\cos(\theta/2) - i\sin(\theta/2) \sigma_i^a \right)\ket{\psi(t_1)}  .
\end{equation}

\paragraph{Time evolution until time $t_2$.}
Time-evolve the rotated system state $\ket{\psi_{\theta}(t_1)}$ up to the time $t_2$,
\begin{equation}
\begin{split}
\ket{\psi_{\theta}(t_2)} &= e^{-iH(t_2-t_1)}\ket{\psi_{\theta}(t_1)} \\
&=\cos(\theta/2) U(t_2)\ket{\psi} - i\sin(\theta/2) U(t_2) \sigma_i^a(t_1) \ket{\psi}  .
\end{split}
\end{equation}

\paragraph{Projective measurement at site $j$.}
Projectively measure observable $\sigma_j^b$, with the probability of measuring eigenvalue $\pm 1$ corresponding to eigenstate $\ket{\pm_b}$ given by Born's rule
\begin{equation}
P^{\text{proj}}_{\pm_b} = \matel{\psi_\theta(t_2)}{\Pi_j^{\pm_b}}{\psi_\theta(t_2)}  .
\end{equation}

\paragraph{Construct expectation value of $\sigma_j^b$.}
Use the above probabilities to construct the expectation value
\begin{multline}\label{e:rotexp}
\matel{\psi_\theta(t_2)}{\sigma_j^b}{\psi_\theta(t_2)} = \cos^2(\theta/2) \expval{\sigma_j^b(t_2)}_{\psi} \\
+ \sin^2(\theta/2) \matel{\psi}{ \sigma_i^a(t_1) \sigma_j^b(t_2) \sigma_i^a(t_1)}{\psi} \\ 
- i \dfrac{1}{2} \sin \theta (-2 i \impart{\matel{\psi}{\sigma_i^a(t_1)  \sigma_j^b(t_2)}{\psi}})  .
\end{multline}
The last term contains the desired imaginary component, while the first two are errors within this context.

\paragraph{Extract $\impart{\matel{\psi}{\sigma_i^a(t_1)  \sigma_j^b(t_2)}{\psi}}$.}
We make use of the fact that the error terms are invariant under a change in parity of the rotation angle $\theta$. By repeating steps \textit{a} to \textit{e} with $\theta \rightarrow - \theta$, and subtracting the new expectation value from \eqref{e:rotexp}, we obtain
\begin{equation}
\begin{split}
&\matel{\psi_\theta(t_2)}{\sigma_j^b}{\psi_\theta(t_2)} - \matel{\psi_{-\theta}(t_2)}{\sigma_j^b}{\psi_{-\theta}(t_2)} \\
&= -2\sin(\theta) \impart{\matel{\psi}{\sigma_i^a(t_1)  \sigma_j^b(t_2)}{\psi}}  .
\end{split}
\end{equation}
The imaginary term is then obtained by inverting the above result
\begin{equation}\label{e:rp_im}
\begin{split}
&\impart{\matel{\psi}{\sigma_i^a(t_1)  \sigma_j^b(t_2)}{\psi}} \\
&= \frac{\matel{\psi_\theta(t_2)}{\sigma_j^b}{\psi_\theta(t_2)} - \matel{\psi_{-\theta}(t_2)}{\sigma_j^b}{\psi_{-\theta}(t_2)}}{-2\sin(\theta)}  .
\end{split}
\end{equation}
We have thus shown that $\impart{\matel{\psi}{\sigma_i^a(t_1)  \sigma_j^b(t_2)}{\psi}}$ can be measured purely by unitary state evolution, followed by a single projective measurement at the final time $t_2$. This protocol suffers no systematic errors, and statistical errors can be minimized by choosing the rotation angle to be $\vert \theta \vert = 3\pi/2$ such that $-2\sin(\theta)=2$.

Combining the protocols of Secs.~\ref{s:DPMP} and \ref{s:RPMP}, we arrive at our second main result: The dynamic correlation function $C(t_1,t_2)=\matel{\psi}{\sigma^a_i(t_1)\sigma_j^b(t_2)}{\psi}$ of an arbitrary spin-$1/2$ model and for arbitrary (in general nonequilibrium) initial states can be measured without the use of ancillas, and with strictly no disturbance due to measurement backaction. This is achieved for the real part of $C$ by a protocol based on projective measurements at times $t_1$ and $t_2$, and for the imaginary part of $C$ by a protocol based on a local rotation at $t_1$ and a projective measurement at $t_2$. From an experimental point of view, ancilla-free measurement schemes are in general much easier to realize, and moreover require only a substantially smaller number of repetitions in order to accumulate sufficient measurement statistics. Systematic errors stemming from a weak coupling expansion are absent, and statistical errors are not amplified, leading to a higher accuracy of the protocol.

\section{Conclusions}

We have presented a theoretical framework for measuring dynamic correlation functions $C(t_1,t_2)$ of arbitrary quantum spin systems, valid in arbitrary equilibrium or nonequilibrium situations. Our first main result, based on Eq.~ \eqref{e:C}, is to show that noninvasive measurements can be used to measure dynamic correlations of general quantum systems. The noninvasive measurement protocol developed in Sec.~\ref{sec:NIM_protocol} uses a weakly coupled ancilla as a noninvasive probe at the earlier time $t_1$. While the use of weakly coupled ancillas is a standard technique to reduce measurement backaction, our main technical result here is that different choices of the system--ancilla coupling operators facilitate the separate measurement of the real and imaginary parts of the dynamic correlation function. In the idealized situation of infinitesimal system--ancilla coupling $\lambda$ and infinite repetitions of the experiment, we show that the exact dynamic correlation function \eqref{e:C_th} is recovered. In the experimentally realistic situation of a finite number $n$ of measurements, statistical as well as systematic errors occur. The error estimates of Sec.~\ref{sec:estimator_construction} and the example of Sec.~\ref{sec:numeric_imp} provide guidance for optimizing experimental parameters.

Our second main result is specific to dynamic correlation functions $C(t_1,t_2)=\matel{\psi}{\sigma_i^a(t_1)\sigma_j^b(t_2)}{\psi}$ of spin-$1/2$ models. In this setting, but for otherwise arbitrary Hamiltonians, we have shown measurement protocols that do not require the use of ancilla degrees of freedom and strictly do not suffer from measurement backaction. The real part of $C$ can be obtained by the protocol of Sec.~\ref{s:DPMP}, which uses projective measurements at times $t_1$ and $t_2$ and correlates the relative frequencies of the outcomes. While the measurement at $t_1$ will influence the state of the system, the correlated outcomes nonetheless yield the correct real part of $C$, with strictly no error due to measurement backaction. The imaginary part of $C$ can be obtained by a protocol based on a local rotation at $t_1$ and a projective measurement at $t_2$, likewise without the use of an ancilla degree of freedom as described in Sec.~\ref{s:RPMP}. Superficially this protocol resembles linear response theory, but, for our spin-$1/2$ setting, is valid to all orders in the rotation angle. These surprising results are valid for arbitrary spin-$1/2$ systems, single-site observables, and initial states (in and out of equilibrium), and greatly facilitate experimental measurements of quantum mechanical dynamic correlations: Ancilla-free measurement schemes are in general much easier to realize, and moreover require a substantially smaller number of repetitions for accumulating sufficient measurement statistics. Systematic errors stemming from a weak coupling expansion are absent, and statistical errors are not amplified, leading to a higher accuracy of the protocol.

Experimental implementations of the measurement protocols should be feasible in a variety of cold atom-based platforms. An experimental scheme for realizing the ancilla-based protocol of Sec.~\ref{sec:NIM_protocol} in linear radio-frequency ion traps was discussed in detail in Sec.~\ref{sec:exp_imp}, concluding that all steps of the noninvasive measurement protocol can be implemented with current technology. The spin-$1/2$ protocols of Secs.~\ref{s:DPMP} and \ref{s:RPMP} do not necessitate the coupling and decoupling of an ancilla, but require single-site resolution and addressability. Experimental realizations should be feasible in quantum gas microscopes, Rydberg-dressed spin lattices \cite{Zeiher_etal16}, and linear ion traps \cite{Kim_etal09,Zhu_etal}.

\begin{acknowledgments}
P.\,U.\ and M.\,K.\ gratefully acknowledge financial support from the Department of Physics and Astronomy of the Universit\"at Heidelberg through the German Excellence Initiative, hospitality at the Kirchhoff-Institut f\"ur Physik, and stimulating discussions with Markus Oberthaler.
P.U.\ acknowledges financial support from the Sam Cohen Trust, the University of Stellenbosch via the Postgraduate Merit Bursary, and the National Research Foundation.
H.U.\ acknowledges financial support by the United States Air Force Office of Scientific Research, Award No.~FA9550-14-1-0151.
M.K.\ acknowledges financial support from the National Research Foundation of South Africa via the Incentive Funding and the Competitive Programme for Rated Researchers.
\end{acknowledgments}

\appendix
\section{Projective measurement protocol: example}
\label{app: naive_prot_eg}
We study the dynamic correlator \eqref{e:C_th} for a lattice consisting of two sites, $i=1$ and $j=2$, with a spin-$1/2$ degree of freedom attached to each of the sites. The dynamics is generated by the Hamiltonian
\begin{equation}
H = (\mvec{n} \cdot \mvec{\sigma})_1 (\mvec{m} \cdot \mvec{\sigma})_2
\end{equation}
with $\mvec{n},\mvec{m}$ being unit vectors.
It follows from the series expansion of the time-evolution operator that
\begin{equation}\label{eq: naive_prot_time_evolution_op}
U(t) = \exp(-iHt)=\cos(t) - i \sin(t) (\mvec{n} \cdot \mvec{\sigma})_1 (\mvec{m} \cdot \mvec{\sigma})_2.
\end{equation}
For simplicity we choose $t_1=0$ and $t_2=t>0$, as well as a product initial state $\ket{\psi} = \ket{\psi_1 \psi_2} = \ket{\psi_1}\otimes \ket{\psi_2}$. In this case one obtains
\begin{multline}\label{eq: naiveProt_exactCF}
C(0,t) = \bra{\psi} \sigma_1^a U^\dagger(t) \sigma_2^b U(t)\ket{\psi}\\ 
= \cos^2 t \bra{\psi_1} \sigma_1^a \ket{\psi_1} \bra{\psi_2}\sigma_2^b \ket{\psi_2} \!+\! \sin^2 t \bra{\psi_1} \sigma_1^a \ket{\psi_1}\\
\times \bra{\psi_2} (\mvec{m} \cdot \mvec{\sigma})_2 \sigma_2^b (\mvec{m} \cdot \mvec{\sigma})_2 \ket{\psi_2} -\frac{i}{2} \sin 2t\\
\times \bra{\psi_1} \sigma_1^a (\mvec{n}\cdot\mvec{\sigma})_1 \ket{\psi_1} \bra{\psi_2} \left[ \sigma_2^b, (\mvec{m} \cdot \mvec{\sigma})_2 \right] \ket{\psi_2}
\end{multline}
for the exact dynamic correlation function \eqref{e:C_th}.

A naive measurement protocol for $C(0,t)$ consists of two projective measurements, one at either time point. The measurement at $t_1=0$ is done in the eigenbasis of $\sigma^a$, which we denote as $\lbrace \ket{+_a}, \ket{-_a} \rbrace$, and the measurement at $t_2=t$ is done in the eigenbasis of $\sigma^b$.

To construct the projective correlation \eqref{e:ps-pd}, we require the probabilities $P_{++}^\text{proj}$, $P_{--}^\text{proj}$, $P_{+-}^\text{proj}$, and $P_{-+}^\text{proj}$. After the first projective measurement at time $t_1=0$, the system state is
\begin{align}\label{e:Psi_pma}
\ket{\psi_{\pm_a}}=\dfrac{\left(\projector{\pm_a} \otimes \mathds{1}\right) \ket{\psi_1 \psi_2}}{\sqrt{P_{\pm_a}^\text{proj}}},
\end{align}
where
\begin{equation}\label{e:Paproj}
P_{\pm_a}^\text{proj} = \bra{\psi_1 \psi_2} \left(\projector{\pm_a}\otimes \mathds{1}\right) \ket{\psi_1 \psi_2} = \absval{\innprod{\pm_a}{\psi_1}}^{2}
\end{equation}
is the probability to measure $+$ or $-$, respectively, in the $\sigma^a$ eigenbasis  \footnote{We have to require an initial state $\protect{\ket{\psi_1}}$ such that $\protect{\innprod{\pm_a}{\psi_1}\neq0}$ in order to avoid division by zero in \eqref{e:Psi_pma}.}. Evolving the system to time $t$ we find the conditional probabilities for measuring site 2 in eigenstate $\ket{\pm_b}$ of $\sigma^b$, given that site 1 was measured in state $\ket{\pm_a}$, to be
\begin{equation}
\begin{split}
P_{\pm_b \vert \pm_a}^\text{proj} &= \matel{\psi_{\pm_a}}{e^{iHt}\left(\Id\otimes\projector{\pm_b}\right)e^{-iHt}}{\psi_{\pm_a}} \\
&= \dfrac{\absval{\innprod{\pm_a}{\psi_1}}^2}{ P_{\pm_a}^\text{proj} } \matel{\pm_a, \psi_2}{e^{iHt}\left(\Id\otimes\projector{\pm_b}\right) \\
&\quad\times e^{-iHt}}{\pm_a, \psi_2} \\
&= \cos^2(t) \absval{\innprod{\pm_b}{\psi_2}}^2 + \sin^2(t) \absval{\matel{\pm_b}{(\mvec{m}\cdot\mvec{\sigma})_2}{\psi_2}}^2 \\
&\quad-\frac{i}{2} \sin(2t) \matel{\pm_a}{(\mvec{n}\cdot\mvec{\sigma})_1}{\pm_a}\\
&\qquad\times \left( \innprod{\psi_2}{\pm_b} \matel{\pm_b}{(\mvec{m}\cdot\mvec{\sigma})_2}{\psi_2} - \text{c.c.}\right).
\end{split}
\end{equation}
Combining these probabilities with \eqref{e:Paproj} according to \eqref{e:ps-pd}, one obtains
\begin{multline}\label{eq: naiveProt_constructedCF}
\mathscr{C}^\text{proj} (0,t) = \Bigl( \cos^2 t \matel{\psi_1}{\sigma^a}{\psi_1} \matel{\psi_2}{\sigma^b}{\psi_2} \\
 + \sin^2 t \matel{\psi_1}{\sigma^a}{\psi_1} \matel{\psi_2}{(\mvec{m} \cdot \mvec{\sigma}) \sigma^b ( \mvec{m} \cdot \mvec{\sigma} ) }{\psi_2}\Bigr) \\
 -i \dfrac{1}{2} \sin(2t) \Bigl( \absval{\innprod{+_a}{\psi_1}}^2 \matel{+_a}{\mvec{n} \cdot \mvec{\sigma}}{+_a}\\
  -  \absval{\innprod{-_a}{\psi_1}}^2  \matel{-_a}{\mvec{n} \cdot \mvec{\sigma}}{-_a} \Bigr) \matel{\psi_2}{ [\sigma^b,  ( \mvec{m} \cdot \mvec{\sigma} ) ] }{\psi_2}.
\end{multline}
Comparing this with \eqref{eq: naiveProt_exactCF} we already see that the naively constructed correlation contains the real part of the dynamic correlation, but the imaginary parts do not match.

To clarify, we substitute the parameters for the simple case mentioned in the introduction, where we considered $zz$ correlations, i.e., $a=b=z$, and a Hamiltonian $H=\sigma_1^x\sigma_2^x$, which corresponds to the choice $\mvec{n}=\mvec{m}=(1,0,0)$. Parametrizing the initial state with respect to the $\sigma^z$ eigenbasis,
\begin{equation}
\ket{\psi_i} = \alpha \ket{+_z} + \beta \ket{-_z}\quad\text{for $i=1,2$,}
\end{equation}
with $\alpha,\beta \in \CC \setminus\{0\}$, the exact dynamic correlation function \eqref{eq: naiveProt_exactCF} reduces to 
\begin{multline}
C(0,t) = \cos(2t) \left( \absval{\alpha}^{2} - \absval{\beta}^{2} \right)^2 \\
-i \sin(2t) \left( \alpha^* \beta - \alpha \beta^* \right)^2,
\end{multline}
whereas \eqref{eq: naiveProt_constructedCF} reduces to
\begin{equation}
\wmcf^\text{proj}(0,t) = \cos(2t) \left( \absval{\alpha}^{2} - \absval{\beta}^{2} \right)^2 = \realpart{C(0,t)}.
\end{equation}

\section{General projective measurement protocol}
\label{app:naive_general}

Here we report details of the derivation of Eq.~\eqref{e:projCond}.
For $s \in \NN/2$, we consider dynamic correlations of spin observables $S_i^a$ and with eigenvalues $m_a \in \mathscr{S}=\lbrace -s, -s+1, \ldots, s-1,s \rbrace$ and corresponding spectral decomposition $S_i^a=\sum_{m_a \in \mathscr{S}} m_a \Pi_i^{m_a}$, where $\Pi_i^{m_a}$ denotes the projector onto the eigenspace corresponding to $m_a$. The projective correlation function \eqref{e:ps-pd} then generalizes to
\begin{equation}
\label{e:ProjCorr}
\mathscr{C}^{\text{proj}}=\sum_{m_a,m_b} m_a m_b P^{\text{proj}}_{m_a}P^{\text{proj}}_{m_b \vert m_a}.
\end{equation}
Upon substituting
\begin{equation}
\label{e:CondProb}
P^{\text{proj}}_{m_b \vert m_a}=\langle \psi | \Pi_i^{m_a}(t_1) \Pi_j^{m_b}(t_2) \Pi_i^{m_a}(t_1) | \psi \rangle / P_{m_a}^\text{proj}
\end{equation}
into (\ref{e:ProjCorr}) we obtain 
\begin{equation}
\label{e:ProjCorr1}
\begin{split}
\mathscr{C}^\text{proj}&=\sum_{m_a,m_b} m_a m_b \langle \psi | \Pi_i^{m_a}(t_1) \Pi_j^{m_b}(t_2) \Pi_i^{m_a}(t_1) | \psi \rangle\\
&=\sum_{m_a} m_a \langle \psi | \Pi_i^{m_a}(t_1) S_j^b(t_2) \Pi_i^{m_a}(t_1) | \psi \rangle .
\end{split}
\end{equation}
Using the identity
\begin{equation}
\Pi_i^{m_a} (t_1)=\mathds{1}_i-\sum_{m'_a \neq m_a} \Pi_i^{m'_a} (t_1)
\end{equation}
to replace the rightmost projector in \eqref{e:ProjCorr1}, we obtain
\begin{align}\label{e:ProjCorr2}
\begin{split}
\mathscr{C}^\text{proj}=&\sum_{m_a} m_a \biggl( \matel{\psi}{ \Pi_i^{m_a} (t_1) S_j^b(t_2)}{\psi} \\
&- \sum_{m'_a \neq m_a} \matel{\psi}{ \Pi_i^{m_a} (t_1) S_j^b(t_2) \Pi_i^{m'_a} (t_1) }{\psi}\biggr)\\
=&\matel{\psi}{S_i^a(t_1) S_j^b(t_2)}{\psi} \\
&- \sum_{m_a} m_a \sum_{m'_a \neq m_a} \matel{\psi}{ \Pi_i^{m_a} (t_1) S_j^b(t_2) \Pi_i^{m'_a} (t_1) }{\psi}.
\end{split}
\end{align}
At this point it is convenient to define the operator
\begin{equation}
\Gamma \equiv \sum_{m_a} m_a \sum_{m'_a \neq m_a} \Pi_i^{m_a} (t_1) S_j^b(t_2) \Pi_i^{m'_a} (t_1),
\end{equation}
with which \eqref{e:ProjCorr2} can be expressed in terms of the exact correlation as
\begin{equation}
\label{e:ProjCorr3}
\mathscr{C}^\text{proj}= C - \matel{\psi}{ \Gamma }{\psi}.
\end{equation}
Since $\matel{\psi}{\left[ S_i^a(t_1),S_j^b(t_2) \right]}{\psi}=2i \impart{C}$ we can write
\begin{equation}
\label{e:ReC}
\realpart{C}=C-\frac{1}{2} \matel{\psi}{ \left[ S_i^a(t_1),S_j^b(t_2) \right] }{\psi}.
\end{equation}
Therefore, equality between (\ref{e:ProjCorr3}) and (\ref{e:ReC}) holds when
\begin{equation}
\label{e:Comm=Gamma}
\matel{\psi}{\left[ S_i^a(t_1),S_j^b(t_2) \right]}{\psi}=2\matel{\psi}{ \Gamma }{\psi}.
\end{equation}
If we express $S_i^a(t_1)$ in the above commutator by its spectral decomposition and introduce the identity $\mathds{1}=\sum_{m'_a} \Pi_i^{m'_a} (t_1)$  at the right of $S_j^b(t_2)$, we find that
\begin{equation}
\label{e:CommProj}
\matel{\psi}{ [ S_i^a(t_1),S_j^b(t_2)]}{\psi}=\matel{\psi}{ \Gamma - \Gamma^{\dagger} }{\psi}.
\end{equation}
Therefore, $\realpart{ C}=\mathscr{C}^\text{proj}$ holds if and only if
\begin{equation}\label{e:gammaCond}
\left( \Gamma - \Gamma^{\dagger} \right) = 2\Gamma,
\end{equation}
i.e., if and only if $\Gamma$ is anti-hermitian.
This shows that validity of $\realpart{C}=\mathscr{C}^\text{proj}$ depends on the spectra of the observables which are to be correlated.

For a spin-$1/2$ system, as in Sec.~\ref{s:DPMP}, we have
\begin{multline}\label{e:Gamma1/2}
2\Gamma_{1/2}=\Pi_i^{+_a} (t_1) S_j^b(t_2) \Pi_i^{-_a}(t_1)\\ - \Pi_i^{-_a} (t_1) S_j^b(t_2) \Pi_i^{+_a}(t_1)
= -2 \Gamma_{1/2}^\dagger  ,
\end{multline}
which satisfies \eqref{e:gammaCond} and thus confirms \eqref{e:dp_real}.

In contrast, for a spin-$1$ system we have $m_a,m_b\in\lbrace 0,\pm1 \rbrace$ and
\begin{multline}
\Gamma_1=\Pi_i^+ (t_1) S_j^b(t_2) \Pi_i^0(t_1) + \Pi_i^+ (t_1) S_j^b(t_2) \Pi_i^-(t_1)\\
- \Pi_i^- (t_1) S_j^b(t_2) \Pi_i^+(t_1)- \Pi_i^- (t_1) S_j^b(t_2) \Pi_i^0(t_1)
\neq -\Gamma_1^\dagger.
\end{multline}
Thus \eqref{e:gammaCond} is not satisfied in general, and one can prove rigorously that $\Gamma$ is anti-hermitian only when the observable to be correlated at $t_1$ has exactly two eigenvalues---which may be degenerate---of the same magnitude, but different sign.

Examples of such observables are single-site spin-$1/2$ observables which are a linear combination of the Pauli matrices, or multi-site spin-$1/2$ observables constructed by taking the tensor product of the aforementioned single-site observables.

\section{Deferred measurement approach}
\label{s:deferred}

In this appendix we show that deferral of the ancilla measurement to times $t\geq t_2$ does not further improve the performance of the noninvasive measurement protocol of Sec.~\ref{sec:NIM_protocol} and gives the same results as the immediate ancilla measurement at time $t_1$.

Up to (and including) step $c.$\ of Sec.~\ref{sec:NIM_protocol}, the protocol remains unchanged. Then, instead of projectively measuring the ancilla state at $t_1$, we keep the post-coupling state $\ket{\Psi_\lambda(t_1)}$ of \eqref{eq: post_weak_coupling_state} unprojected, and proceed by time-evolving that state with the system Hamiltonian $H_\text{S}$ until time $t_2$,
\begin{equation}
	\ket{\Psi(t_2)} \simeq \Id_\text{A} \otimes \,e^{-iH_\text{S}(t_2-t_1)} \ket{\Psi(t_1)}.
\end{equation}
The joint probabilities for the different combinations of measured outcomes are obtained by calculating
\begin{equation}\label{eq: def_meas_probability}
	P_{m_a m_b} \simeq \matel{\Psi(t_2)}{ \left( \projector{m_a} \otimes \projector{m_b} \right) }{\Psi(t_2)},
\end{equation}
where the $m_a$--projector acts on the ancilla, and the $m_b$--projector only on site $j$ of the system. Combining these probabilities according to \eqref{eq: wmcf} we obtain
\begin{equation}
\begin{split}
	\wmcf =& \sum_{m_a,m_b \in \mathscr{s}}m_a m_b P_{m_a m_b}\\
	=& \matel{\Psi(t_2)}{  \sum_{m_a} m_a\projector{m_a}  \otimes \sum_{m_b} m_b \projector{m_b}}{\Psi(t_2)}\\
	=& \matel{\Psi(t_2)}{ S^a S_j^b }{\Psi(t_2)}\\
	\simeq& \expval{S^a}_{\phi} \expval{S_j^b(t_2)}_{\psi}\\
	&- i \lambda \left[ \expval{S^a B}_{\phi} \matel{\psi}{S_j^b(t_2) A_i(t_1) }{\psi} - \text{ c.c.}  \right].
\end{split}
\end{equation}
Expressing $\ket{\phi}$ in the eigenbasis of $S^a$ as in \eqref{e:phi}, Eq.~\eqref{e:16} is reproduced, which confirms that immediate and deferred ancilla measurements give identical results to leading order in $\lambda$. A more general calculation reveals that the two approaches are equivalent to all orders in $\lambda$.

From a theoretical point of view deferred measurements have the advantage that no linearization of the post-ancilla-measurement system state (as in \eqref{e:postMeasLin}) is required. We will exploit this advantage when deriving a protocol that involves multiple noninvasive measurements in Appendix~\ref{app: Calculations_for_CMP}. Experimentally the advantage of one or the other protocol is less clear. One may imagine experimental platforms in which storing the ancilla state until later times is difficult (favoring immediate measurement), or other situations in which the immediate measurement of the ancilla generates unwanted noise (favoring deferred measurement).

\section{Multiple measurements}
\label{sec:multiple_times}

The protocol of Sec.~\ref{sec:NIM_protocol} describes a procedure to noninvasively measure a dynamic correlation function $C(t_1,t_2)$ at a fixed pair of times $(t_1,t_2)$. In physical applications, one will frequently be interested in more than one such pair, or even in the functional dependence of $C$ over a range of times. In this section we will investigate and compare different strategies for noninvasively measuring dynamic correlation functions in that situation.

To keep the discussion simple, we consider a minimal model consisting of two spin-$1/2$ degrees of freedom, and focus on dynamic correlation functions
\begin{subequations}
\begin{align}
C(t_1,t_2) &= \matel{\psi}{ \sigma_1^a (t_1) \sigma_2^b (t_2)}{\psi}, \label{eq: mult_corr_a}\\
C(t_1,t_3) &= \matel{\psi}{ \sigma_1^a (t_1) \sigma_2^b (t_3)}{\psi}, \\
C(t_2,t_3) &= \matel{\psi}{ \sigma_2^b (t_2) \sigma_1^a (t_3)}{\psi}, \label{eq: mult_corr_c}
\end{align}
\end{subequations}
at three points in time, $t_3>t_2>t_1\geq0$.
One obvious way of noninvasively measuring these correlations is by repeating the protocol of Sec.~\ref{sec:NIM_protocol} separately for each of the three correlations \eqref{eq: mult_corr_a}--\eqref{eq: mult_corr_c}. We refer to this procedure as the single-noninvasive measurement protocol (sNIMP), as it involves only one noninvasive measurement before the final projective one.

In an attempt to avoid multiple, possibly very large, data samples one might hope to develop a more efficient protocol based on noninvasive measurements at $t_1$ and $t_2$, followed by a projective measurement at $t_3$, during each repetition of the experiment. We will refer to this protocol as the consecutive-noninvasive measurement protocol (cNIMP). While both protocols turn out to be feasible in principle, they differ in their efficiency. Here we assume that, like in many experiments, the number of repetitions of the experiment is a limiting factor, and we will investigate in the following whether the sNIMP or the cNIMP  is more efficient at determining all three correlations \eqref{eq: mult_corr_a}--\eqref{eq: mult_corr_c} to a desired accuracy. 

To implement the cNIMP we need two ancilla spins, one coupled to site $1$ at $t_1$ with coupling time $\lambda_1$, the other to site $2$ at $t_2$ with coupling time $\lambda_2$. This allows us to measure $C_{n}^\lambda(t_1,t_2)$, while simultaneous projective measurements of sites $1$ and $2$ at $t_3$ allow us to measure $C_{n}^\lambda (t_2,t_3)$ and $C_{n}^\lambda (t_1,t_3)$, respectively. The derivation of the \mbox{cNIMP} estimators, the required coupling operators, and associated errors is similar to that of Sec.~\ref{sec:NIM_protocol} and can be found in Appendix~\ref{app: Calculations_for_CMP}. Most importantly, the cNIMP has to be executed only thrice to obtain all six estimators of the real and imaginary parts of correlations \eqref{eq: mult_corr_a}--\eqref{eq: mult_corr_c}.
Other choices of correlation functions than those in Eqs.~\eqref{eq: mult_corr_a}--\eqref{eq: mult_corr_c} may require more than two ancillas, but derivations go along similar lines.

Since $C_{n}^\lambda (t_1,t_2)$ is obtained from two consecutive noninvasive measurements, its estimator \eqref{eq:cNIMP_estt1t2} involves a division by both coupling parameters, $\lambda_1$ and $\lambda_2$. As a consequence, the associated statistical error will be amplified much stronger than in the sNIMP. Pushing this error below a certain desired level therefore requires large sample sizes $n$, as shown in Fig.~\ref{fig: minEstDev_vs_N_CMPvsSMP}, and is the reason for the inferior performance of the cNIMP. The estimators $C_{n}^\lambda(t_1,t_3)$ and $C_{n}^\lambda(t_2,t_3)$ as given in \eqref{eq:cNIMP_estt1t3} and \eqref{eq:cNIMP_estt2t3} involve a division by only one of the coupling parameters $\lambda_1$ or $\lambda_2$, and so the resulting total errors, while still greater than in the sNIMP, are at least of the same order of magnitude (see Fig.~\ref{fig: minEstDev_vs_N_CMPvsSMP}, right plot).
The reason why the errors for these two correlations are still larger is due to a larger systematic error which is incurred for non-zero $\lambda_1$ and $\lambda_2$.
 
\begin{figure}\centering
  \includegraphics[width=0.49\linewidth]{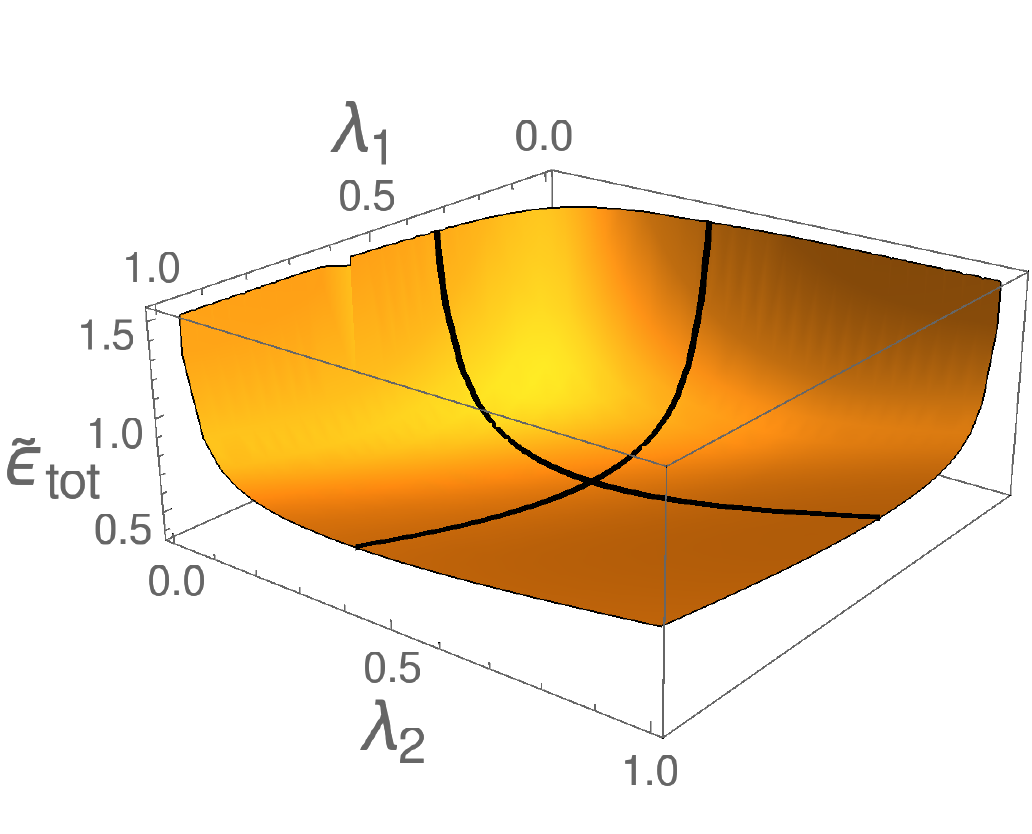}
  \hfill
  \includegraphics[width=0.49\linewidth]{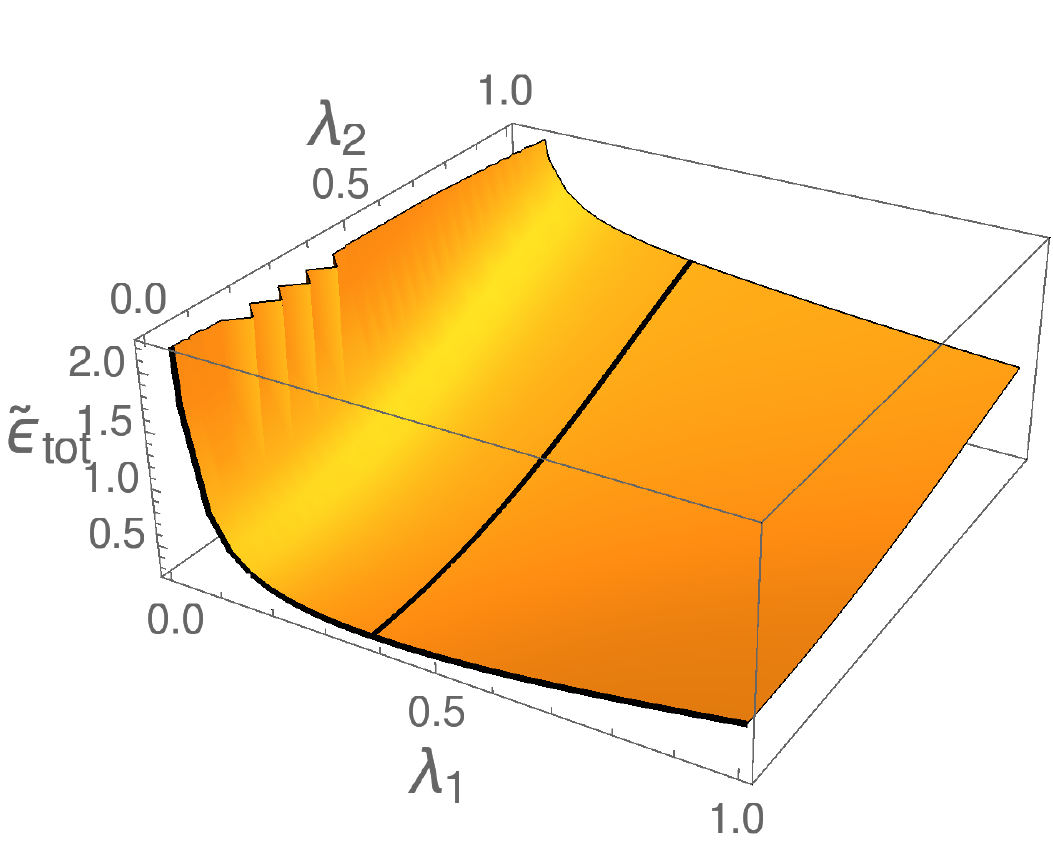}
\caption{\label{fig: boundsCMP}%
Predicted upper bounds on the relative error $\tilde{\epsilon}_\text{tot}$ for measurements of correlations $C(t_i,t_j)$ within the cNIMP for times $(t_1,t_2,t_3)=(0,1,10)$, initial system state parameters as in Fig.~\ref{fig: predErr_manySamples}, and sample size $n=10^5$.
Left: Estimated total relative error for measurements of $C_{n}^\lambda(t_1,t_2)$. The black curves are included to guide the reader's eye, and their intersection indicates the minimum error of $37\%$ at $(\lambda_1^{*},\lambda_2^{*})=(0.40,0.41)$.
Right: Corresponding prediction for measurements of $C_{n}^\lambda(t_1,t_3)$, exhibiting a minimum error of $25\%$ at $(\lambda_1^{*},\lambda_2^{*})=(0.37,0.00)$. Although measurement of this correlation is performed in the cNIMP by observing states of only the first ancilla (coupled to lattice-site $1$ at $t_1$) and the spin at site 2, the additional coupling of the second ancilla to site $2$ at intermediate time $t_2$ increases the systematic error, which causes the total error to increase with $\lambda_2$. This reflects, and is due to, the fact that the cNIMP perturbs the system's dynamics more strongly.
}%
\end{figure}

To illustrate the discussed findings, and compare this protocol to the sNIMP, we revisit the example of Sec.~\ref{sec:numeric_imp} with Ising-type Hamiltonian \eqref{e:Hxx} and $zz$ correlation functions, i.e., $a=b=z$ in \eqref{eq: mult_corr_a}--\eqref{eq: mult_corr_c}. 
Figure \ref{fig: boundsCMP} shows the estimated total error $\tilde{\epsilon}_\text{tot}$ for the cNIMP as a function of both coupling times $\lambda_1$ and $\lambda_2$ for $C_{n}^\lambda(t_1,t_2)$ and $C_{n}^\lambda(t_1,t_3)$.
For $C_{n}^\lambda(t_1,t_2)$ a clear minimum deviation of $37\%$ is indicated by the intersection of the black curves at $(\lambda_1^{*},\lambda_2^{*})=(0.40,0.41)$.
Beyond this optimal coupling coordinate, the accuracy of the cNIMP estimator $C_{n}^\lambda(t_1,t_2)$ is bad as the total deviation grows to be on the order of $\vert C(t_1,t_2) \vert$. 
In the regime where either coupling time is small this large deviation is due to the above mentioned amplification of the statistical error brought about by the $1/(\lambda_1 \lambda_2)$ factor in \eqref{eq:cNIMP_estt1t2}.
For larger coupling times, systematic errors incurred with respect to both coupling parameters add up to yield a larger systematic error than in the sNIMP.

The estimator $C_{n}^\lambda(t_1,t_3)$ is obtained in the cNIMP from measurements of the first ancilla at $t_1$ and of site $2$ at $t_3$. At the intermediate time $t_2$ the cNIMP perturbs the system dynamics by coupling a second ancilla to site $2$. This perturbation is reflected in the error bound of $C_{n}^\lambda(t_1,t_3)$ (Fig.~\ref{fig: boundsCMP}, right) which increases also with the coupling time $\lambda_2$.
We omit the error bound of $C_{n}^\lambda(t_2,t_3)$ as it reflects a similar behaviour, only with the roles of $\lambda_1$ and $\lambda_2$ interchanged.

To measure \eqref{eq: mult_corr_a}--\eqref{eq: mult_corr_c} with accuracies as in Fig.~\ref{fig: boundsCMP} one needs a total of three samples of $n=10^5$ measurements. In Fig.~\ref{fig: predicted_errors} we showed that the sNIMP achieves similar accuracies for samples of $n=10^4$ measurements per real and imaginary component. This is a first indication that the cNIMP is less efficient than the sNIMP due to its lower accuracy, which we attribute to the repeated perturbation of the system dynamics.

\begin{figure}\centering
  \includegraphics[width=0.49\linewidth]{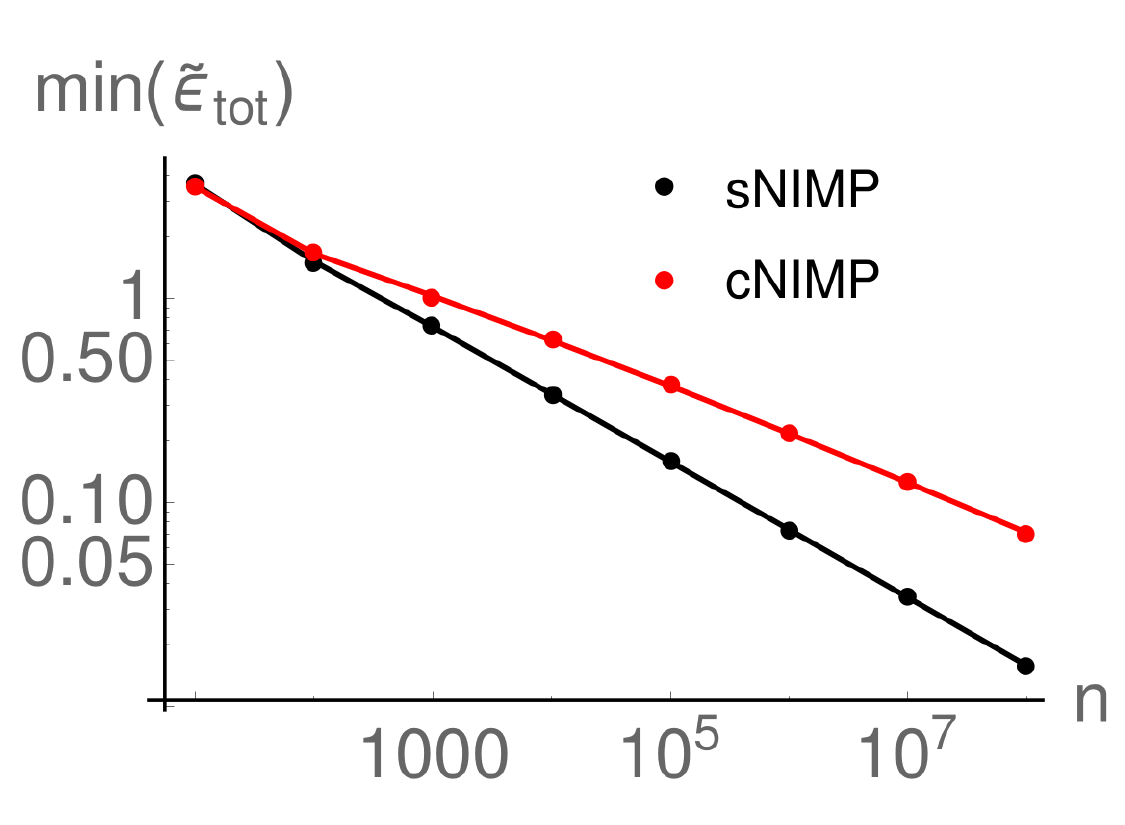}
  \hfill
  \includegraphics[width=0.49\linewidth]{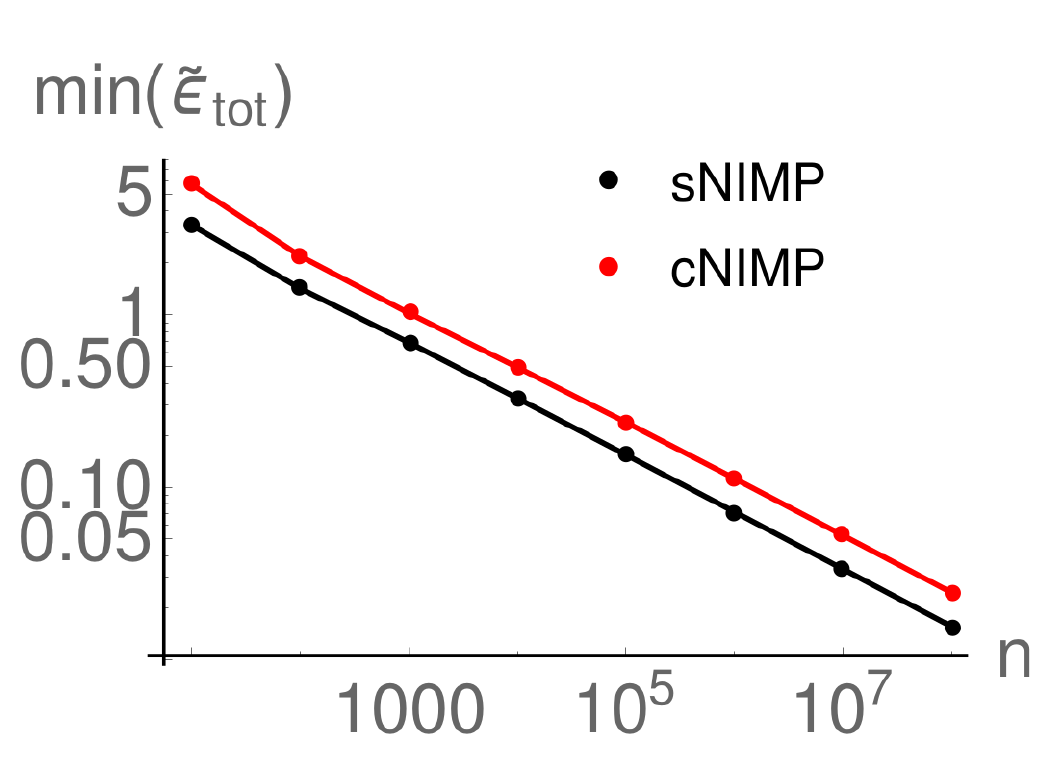}
\caption{\label{fig: minEstDev_vs_N_CMPvsSMP}%
Minimum total error $\min_\lambda \tilde{\epsilon}_{\text{tot}}(\lambda)$ as a function of sample size $n$. Times and system parameters are as in Fig.~\ref{fig: boundsCMP}. Left: For $C_{n}^\lambda(t_1,t_2)$ the minimum error of the sNIMP (lower black line) decreases faster than that of the cNIMP (upper red line). As a result, to construct an estimator with a total error of $10\%$ or less, sample sizes in the sNIMP must be at least $10^6$, which is about two orders of magnitude smaller than for the cNIMP. 
Right: For measurements of $C_{n}^\lambda(t_1,t_3)$ the minimum error of either protocol decreases at the same rate with the sample size, however the errors of the cNIMP (upper red line) are consistently larger than those of the sNIMP (lower black line).
Results for $C_{n}^\lambda(t_2,t_3)$ are similar (not shown).
}%
\end{figure}
To test this expectation we calculated, for both protocols, the minima of the predicted estimator deviation $\tilde{\epsilon}_{\text{tot}}$ for increasing sample sizes (Fig.~\ref{fig: minEstDev_vs_N_CMPvsSMP}). Especially for the estimator $C_{n}^\lambda(t_1,t_2)$, which requires two noninvasive measurements in the cNIMP, the sNIMP is much more efficient in the large $n$ regime (where the bound \eqref{e:estat_bound} is valid).
The plot shows that for this correlation, the sNIMP error decreases at a faster rate than the cNIMP error such that an accuracy of $10\%$ or less can be achieved in the sNIMP from $2 \times 10^6$ measurements, while in the cNIMP one would require $2 \times 10^8$ measurements.
The minimum errors for the other two estimators decrease at the same rate in both protocols, but are consistently smaller in the sNIMP.

In summary, to measure correlations \eqref{eq: mult_corr_a}-\eqref{eq: mult_corr_c} with an accuracy of at least $10\%$, the cNIMP and sNIMP require, respectively, a net sample size $n_c = 3 \times 10^8$ and $n_s = 6 \times 10^6$.
This example shows that multiple dynamic correlations are most efficiently measured with repeated implementations of the sNIMP.

\section{Derivation of the cNIMP}
\label{app: Calculations_for_CMP}

As outlined in Appendix~\ref{sec:multiple_times}, the consecutive-noninvasive measurement protocol (cNIMP) consists of two noninvasive measurements at times $t_1$ and $t_2$ (one at either time), followed by a projective measurement at $t_3$. To keep the calculations simple, we use the deferred measurement approach, which in Appendix~\ref{s:deferred} was shown to give the same results as immediate measurements of the ancilla spins. For notational simplicity we derive the results in the language of spin-$1/2$ models, but generalizations to $s>1/2$ are straightforward.

To perform two noninvasive measurements we require two ancilla spins. The total Hilbert space is therefore $\mathscr{H}=\mathscr{H}_{\text{A}_1} \otimes \mathscr{H}_{\text{A}_2} \otimes \mathscr{H}_\text{S}$, where $\mathscr{H}_{\text{A}_m}=\CC^2$ denotes the Hilbert space of ancilla $m$. As an initial state we use $\ket{\Psi}=\ket{\phi_1,\phi_2,\psi}$, with ancilla initial states $\ket{\phi_m}$ to be determined.

The relevant time evolution operators on $\mathscr{H}$ for the protocol are
\begin{subequations}
\begin{align}
U(t) &= \mathds{1} \otimes \mathds{1} \otimes \exp(-iH_\text{S}t),\\
\couple (\lambda_1) &= \exp(-i \lambda_1 B_1 \otimes \mathds{1}_{A_2} \otimes A_i),\\
\couple (\lambda_2) &= \exp(-i \lambda_2 \mathds{1}_{A_1} \otimes B_2 \otimes A_j),
\end{align}
\end{subequations}
which describe the system dynamics, the coupling to the first ancilla, and the coupling to the second ancilla, respectively. The coupling operators $A_i,A_j$ act nontrivially only on lattice sites $i$ and $j$, respectively. In terms of the above time evolution operators, the state at time $t_3$ is given by
\begin{equation}\label{eq: CMP_initial_combined_state}
\ket{\Psi (t_3)} = U(t_3-t_2) \couple (\lambda_2) U(t_2-t_1) \couple (\lambda_1)U(t_1) \ket{\Psi}.
\end{equation}

Using the deferred measurement approach to measure \eqref{eq: mult_corr_a}, at time $t_3$ ancilla 1 is measured in the eigenbasis $\lbrace \ket{+_a},\ket{-_a} \rbrace$ of $\sigma^a$, and ancilla 2 is measured in the eigenbasis of $\sigma^b$. The joint probabilities for these measurements are then
\begin{align}
P_{\pm_a \pm_b} & = \matel{\Psi(t_3)}{\projector{\pm_a} \otimes \projector{\pm_b} \otimes \mathds{1}_\text{S} }{\Psi(t_3)}.
\end{align}
Next we combine the probabilities of the four combinations of outcomes as in \eqref{eq: wmcf}. For the choices $A_1=\sigma^a$, $A_2=\sigma^b$, $\ket{\phi_1}=(\ket{+_a}+\ket{-_a})/\sqrt{2}$ and $\ket{\phi_2}=(\ket{+_b}+\ket{-_b})/\sqrt{2}$ one obtains, to leading order in the couplings $\lambda_1$ and $\lambda_2$, 
\begin{multline}
\wmcf(t_1,t_2) = \lambda_1 \lambda_2 \left( \matel{-_b}{B_2}{+_b} - \matel{+_b}{B_2}{-_b} \right)\\
\times\Bigl[ \left( \matel{+_a}{ B_1}{-_a} - \matel{-_a}{B_1}{+_a} \right) \realpart{C(t_1,t_2)}\\
-i\left( \expval{B_1}_{+_a} -\expval{B_1}_{-_a} \right) \impart{C(t_1,t_2)} \Bigr],
\end{multline}
from which one can read off that $B_2=i\ket{-_b}\bra{+_b}-i\ket{+_b}\bra{-_b}$ is a suitable choice to maximize the prefactor on the right-hand side of this equation. 
Similar to the sNIMP protocol of Sec.~\ref{sec:NIM_protocol}, imaginary and real parts of $C(t_1,t_2)$ are obtained by using $B_1=B^{(1)}=\sigma^a$ and $B_1=B^{(2)}=i\ket{-_a}\bra{+_a}-i\ket{+_a}\bra{-_a}$, respectively. Taking all of this together, we can construct the estimator
\begin{equation}\label{eq:cNIMP_estt1t2}
C^\lambda(t_1,t_2) = \frac{\wmcf^{(2)}(t_1,t_2) + i\wmcf^{(1)}(t_1,t_2)}{4\lambda_1\lambda_2},
\end{equation}
where the superscripts indicate whether $B^{(1)}$ or $B^{(2)}$ has been used for $B_1$ in the system--ancilla coupling.
We find by similar calculations that the estimator of $C(t_1,t_3)$ is obtained with $B_2$ chosen such that $\expval{B_2}_{\phi_2} = 0$. Due to the above restrictions on the two initial ancilla states there are then two suitable choices of $B_2$ 
\begin{subequations}\label{eq:B2_estt1t3}
\begin{align}
B_2 &=  \ket{+_b}\bra{+_b} -\ket{-_b}\bra{-_b} ,\\ 
B_2 &= i \ket{-_b}\bra{+_b} - i \ket{+_b}\bra{-_b} .
\end{align}
\end{subequations}
Estimators of imaginary and real components are obtained with the same choices of $B_1$ as for \eqref{eq:cNIMP_estt1t2} and so
\begin{align}\label{eq:cNIMP_estt1t3}
C^\lambda(t_1,t_3) &= - \frac{\wmcf^{(2)}(t_1,t_3) + i\wmcf^{(1)}(t_1,t_3)}{2\lambda_1} .
\end{align}
Whereas the above estimator is obtained from the first weak measurement, the estimator of $C(t_2,t_3)$ is obtained from the second. Therefore, the roles of $B_1$ and $B_2$ are reversed and $C^\lambda(t_2,t_3)$ is obtained with
\begin{subequations}\label{eq:B1_estt2t3}
\begin{align}
B_1 =  \ket{+_a}\bra{+_a} -\ket{-_a}\bra{-_a} \text{ or}\\ 
B_1 = i \ket{-_a}\bra{+_a} - i \ket{+_a}\bra{-_a} .
\end{align}
\end{subequations}
Estimators for the imaginary and real parts then require $B_2 = B^{(3)} = \ket{+_b}\bra{+_b} -\ket{-_b}\bra{-_b}$ and $B_2= B^{(4)} = i \ket{-_b}\bra{+_b} - i \ket{+_b}\bra{-_b}$ respectively, so that
\begin{align}\label{eq:cNIMP_estt2t3}
C^\lambda(t_2,t_3) &= - \frac{\wmcf^{(4)}(t_2,t_3) + i\wmcf^{(3)}(t_2,t_3)}{2\lambda_2} .
\end{align}

To summarize, the cNIMP requires
\begin{enumerate}
\item $A_1=\sigma_1^a$ and $A_2=\sigma_2^b$, and $\ket{\phi_1}=(\ket{+a}+\ket{-a})/\sqrt{2}$ and $\ket{\phi_2}=(\ket{+b}+\ket{-b})/\sqrt{2}$  for all measurements.
\item $C^\lambda(t_1,t_2)$: $B_2=i\ket{-_b}\bra{+_b}-i\ket{+_b}\bra{-_b}$ for both components and
\begin{itemize}
\item $B_1=B^{(1)}=\sigma^a$ for the imaginary component,
\item $B_1=B^{(2)}=i\ket{-_a}\bra{+_a}-i\ket{+_a}\bra{-_a}$ for the real component.
\end{itemize}
\item $C^\lambda(t_1,t_3)$: two choices of $B_2$ \eqref{eq:B2_estt1t3} and,
\begin{itemize}
\item $B_1=B^{(1)}=\sigma^a$ for the imaginary component,
\item $B_1=B^{(2)}=i\ket{-_a}\bra{+_a}-i\ket{+_a}\bra{-_a}$ for the real component.
\end{itemize}
\item $C^\lambda(t_2,t_3)$: two choices of $B_1$ \eqref{eq:B1_estt2t3} and,
\begin{itemize}
\item $B_2=B^{(3)}=\sigma^b$ for the imaginary component,
\item $B_2=B^{(4)}=i\ket{-_b}\bra{+_b}-i\ket{+_b}\bra{-_b}$ for the real component.
\end{itemize}
\end{enumerate}
Due to the flexibility of $B_1$ and $B_2$ for estimators $C^\lambda(t_1,t_3)$ and $C^\lambda(t_2,t_3)$, one can measure multiple estimators of real and imaginary parts simultaneously, which is a potential advantage of the cNIMP over the sNIMP.

For the example of Appendix~\ref{sec:multiple_times} where $a=b=z$, we can measure all 6 components with 3 iterations of the cNIMP as follows:
\begin{enumerate}
\item $B_1=\sigma^z$, $B_2= \sigma^y$: \\ $\wmcf^{(1)}(t_1,t_2)$ and $\wmcf^{(1)}(t_1,t_3)$,
\item $B_1=\sigma^y$, $B_2 = \sigma^y$: \\  $\wmcf^{(2)}(t_1,t_2)$ and $\wmcf^{(2)}(t_1,t_3)$ and $\wmcf^{(4)}(t_2,t_3)$,
\item $B_1=\sigma^y$, $B_2 = \sigma^z$: \\ $\wmcf^{(3)}(t_2,t_3)$.
\end{enumerate}
The fact that we can measure all 6 components from only 3 samples allows the cNIMP to potentially be more efficient than the sNIMP.
Statistical errors of the estimators are calculated in the same manner as for the sNIMP.

\section{Ancilla and system rotations for TPM coupling}
\label{app:ancillaRot}

When using the transverse phonon mode (TPM) coupling described in Sec.~\ref{sec:exp_imp}, a coupling Hamiltonian of type $H_c =B \otimes A_i = \sigma^z \otimes \sigma_i^z$ is induced. The noninvasive measurement protocol of Sec.~\ref{sec:NIM_protocol} requires more flexibility in order to obtain estimators of the real and imaginary parts of \eqref{e:C_th} as outlined in \eqref{e:16}--\eqref{e:C}. By augmenting the TPM coupling with rotations of the ancilla and system spins, we show that all the required types of coupling Hamiltonians can be implemented, allowing one to measure dynamic correlations with any combination of $a,b \in \lbrace x,y,z \rbrace$. To simplify the presentation we use the deferred measurement approach of Appendix \ref{s:deferred}.

The overall ancilla-system state at $t_2$ is then
\begin{multline}\label{eq:rotState}
\ket{\Psi_R(t_2)} = U(t_2-t_1) (R_A(\alpha, \mvec{m}) R_S(\theta, \mvec{n}))^\dagger \couple(\lambda) \\
\times R_A(\alpha, \mvec{m}) R_S(\theta, \mvec{n}) \ket{\phi, \psi }
\end{multline}
where the rotations of the system and ancilla are respectively 
\begin{align}
 R_S(\theta, \mvec{n}) &= \prod_{k=1}^{N}  R_k(\theta, \mvec{n})   = \prod_{k=1}^{N} \exp \left(- \dfrac{i\theta}{2} (\mvec{n} \cdot \mvec{\sigma})_k \right), \\
 R_A(\alpha, \mvec{m}) &= \exp \left(-i \dfrac{\alpha}{2} (\mvec{m} \cdot \mvec{\sigma}) \right).
\end{align}
Expanding \eqref{eq:rotState} and keeping $B \otimes A_i$ general for now, we obtain
\begin{multline}
\ket{\Psi_R(t_2)}\\ = \ket{\phi, \psi(t_2)} - \lambda B(\alpha) \ket{\phi} \otimes U(t_2-t_1) A_i(\theta) \ket{\psi(t_1)},
\end{multline}
where $B(\alpha) = R_A^\dagger(\alpha, \mvec{m}) B R_A(\alpha, \mvec{m})$ and $A_i(\theta)  = R_i^\dagger(\theta, \mvec{n}) A_i R_i(\theta, \mvec{n})$.
From a theoretical point of view, a local rotation of only the $i$th spin yields the same state as above. The decision of whether to perform a global or a local rotation of the system is then one which depends on the experimental set up at hand.

Combining probabilities $P_{\pm_a \pm_b}$ as in \eqref{eq: wmcf} we get
\begin{multline}
\wmcf(t_1,t_2) \simeq \expval{\sigma^a}_\phi \expval{\sigma_j^b(t_2)}_\psi \\
- i \lambda \Bigl( \matel{\phi}{\sigma^a B(\alpha)}{\phi} \matel{\psi}{\sigma_j^b(t_1) A_i(\theta, t_1)}{\psi} - \text{ c.c.} \Bigr),
\end{multline}
where $A_i(\theta, t_1) = U^\dagger(t_1) A_i(\theta) U(t_1)$.

Recalling that $B \otimes A_i = \sigma^z \otimes \sigma_i^z$, the estimators \eqref{e:CB1} and \eqref{e:CB2} can then be obtained if the ancilla rotation $R_A(\alpha,\mvec{m})$ is chosen such that $B(\alpha)$ satisfies conditions \eqref{e:B1} and \eqref{e:B2}, respectively.

For $a=x,y$, the above is achieved when the system rotation axis is orthogonal to the $az$ plane, while for $a=z$ no system rotation is needed since $A_i = \sigma_i^z$ is already fulfilled by the TPM coupling.
The same is true for the ancilla rotation when measuring \eqref{e:CB1} with $a=x,y$ whereas for $a=z$ no rotation is needed since $B=\sigma^z$ already fulfills condition \eqref{e:B1}. 
Estimators \eqref{e:CB2} are obtained for $a=x,y$ by rotating the ancilla parallel to the $a$-axis, while for $a=z$ a rotation around the $x$ axis is necessary.
A summary of the appropriate rotations is given in Table \ref{t:rot}. 
\begin{table}[!h]
\centering
\caption{Summary of rotation parameters needed to measure components of $\matel{\psi}{\sigma_i^a(t_1) \sigma_j^b(t_2)}{\psi}$}
\label{t:rot}
\begin{tabular}{|l|l|l|l|l|l|l|l|}
\hline
$a$                  & \text{Component} & $\mvec{n}$                  & $\theta$                  & $A_i(\theta)$                 & $\mvec{m}$ & $\alpha$ & $B$ \\ \hline
\multirow{2}{*}{$x$} & $\realpart{\expval{\sigma_i^x(t_1)\sigma_j^b(t_2)}}$    & \multirow{2}{*}{$(0,1,0)$} & \multirow{2}{*}{$3\pi/2$} & \multirow{2}{*}{$\sigma_i^x$} & $(1,0,0)$ & $\pi/2$  & $\sigma^y$  \\ \cline{2-2} \cline{6-8} 
                     & $\impart{\expval{\sigma_i^x(t_1)\sigma_j^b(t_2)}}$     &                            &                           &                             & $(0,1,0)$ & $3\pi/2$ & $\sigma^x$  \\ \hline
\multirow{2}{*}{$y$} & $\realpart{\expval{\sigma_i^y(t_1)\sigma_j^b(t_2)}}$     & \multirow{2}{*}{$(1,0,0)$} & \multirow{2}{*}{$\pi/2$}  & \multirow{2}{*}{$\sigma_i^y$} & $(0,1,0)$ & $3\pi/2$ & $\sigma^x$  \\ \cline{2-2} \cline{6-8} 
                     & $\impart{\expval{\sigma_i^y(t_1)\sigma_j^b(t_2)}}$    &                            &                           &                             & $(1,0,0)$ & $\pi/2$  & $\sigma^y$  \\ \hline
\multirow{2}{*}{$z$} & $\realpart{\expval{\sigma_i^z(t_1)\sigma_j^b(t_2)}}$    & \multirow{2}{*}{}          & \multirow{2}{*}{$0$}      & \multirow{2}{*}{$\sigma_i^z$} & $(1,0,0)$ & $\pi/2$  & $\sigma^y$  \\ \cline{2-2} \cline{6-8} 
                     & $\impart{\expval{\sigma_i^z(t_1)\sigma_j^b(t_2)}}$    &                            &                           &                             &           & $0$      & $\sigma^z$  \\ \hline
\end{tabular}
\end{table}

\bibliography{LRLR}

\end{document}